\documentclass[prc,reprint,tightenlines,superscriptaddress,nofootinbib,floatfix,longbibliography]{revtex4-2}
\usepackage{amsmath}
\usepackage{amssymb}
\usepackage{bm}
\usepackage{graphicx}
\usepackage{url} 
\usepackage{ulem}
\usepackage{hyperref}

\newcommand{\ket}[1]{|{#1}\rangle}
\newcommand{\bra}[1]{\langle{#1}|}
\newcommand{\inp}[2]{\langle{#1}|{#2}\rangle}
\newcommand{\cg}[3]{({#1},{#2}|{#3})}

\newcommand{\moms}[1]{\bm{#1}}

\newcommand{\mi}{\pi}
\newcommand{\bi}{N}
\newcommand{\pex}{{\bar K_{\rm ex}}}
\newcommand{\vex}{{\bar K^{*}_{\rm ex}}}
\newcommand{\mex}{{M_{\rm ex}}}
\newcommand{\mf}{{K^*}}
\newcommand{\mfo}{{\pi_\mathrm{f}}}
\newcommand{\mft}{{K_\mathrm{f}}}
\newcommand{\mx}{M}
\newcommand{\bx}{B}

\newcommand{\mmi}{m_\mi}
\newcommand{\mbi}{m_\bi}

\newcommand{\mmex}{m_{\mex}}
\newcommand{\mmf}{M_{\mf}}
\newcommand{\mmfo}{m_{\mfo}}
\newcommand{\mmft}{m_{\mft}}
\newcommand{\mmx}{m_M}
\newcommand{\mbx}{m_B}

\newcommand{\lfr}{\mathrm{L}}

\begin{document}

\title{Low-lying $\Lambda$ and $\Sigma$ resonances studied with the forward $K^*$ productions off the proton induced by a high-momentum $\pi$ beam}

\author{H. Kamano}
\email{kamano@rcnp.osaka-u.ac.jp}
\affiliation{Research Center for Nuclear Physics, The University of Osaka, Ibaraki, Osaka 567-0047, Japan}

\author{T.-S. H. Lee}
\email{tshlee@anl.gov}
\affiliation{Physics Division, Argonne National Laboratory, Argonne, Illinois 60439, USA}

\date{\today}

\begin{abstract}
We develop a novel model utilizing the forward $K^*$ production reaction off the nucleon, $\mi \bi \to \mf \mx \bx$, induced by a high-momentum $\pi$ beam, 
as a tool to study low-lying $Y^*$ resonances below and just above the $\bar{K}N$ threshold. 
Because conventional $K^- p$ scattering experiments face difficulties in directly accessing this kinematic region, 
the proposed reaction offers a valuable complementary approach for $Y^*$ spectroscopy. 
The constructed model is based on the one-meson exchange mechanism, 
which is known to dominate forward-angle production at high energies, 
and the half-off-shell scattering amplitudes from the ANL-Osaka dynamical coupled-channels models (Model~A and Model~B). 
We predict various observables, including differential cross sections and angular distributions. 
Our results demonstrate significant enhancements in the subthreshold region of the invariant mass spectra. 
Notably, we show that overlapping resonances, such as a potential new $3/2^+$ $\Sigma$ state and the well-established $\Sigma(1385)3/2^+$, 
can constitute a single peak in the $\pi\Lambda$ mass spectrum, 
indicating that the existence of previously unconfirmed subthreshold states cannot be ruled out by analyzing only the existing mass spectrum data. 
Furthermore, we find that angular distributions provide strong discriminatory power to disentangle such overlapping states through partial-wave interference effects, 
while the $t'$ and $\phi_\mx^*$ dependencies provide crucial constraints on the high-energy production mechanisms.
Our predictions for these highly sensitive observables can facilitate high-statistics measurements, 
which are accessible at modern hadron facilities such as J-PARC, 
to unravel the $S=-1$ $Y^*$ mass spectrum.
\end{abstract}

\maketitle

\section{\label{sec:intro}Introduction}

Significant progress has been made in the last decade regarding the spectroscopy of 
$\Lambda$ and $\Sigma$ hyperon resonances with strangeness $S=-1$ (collectively referred to as $Y^*$).
A key factor in this development has been the use of advanced coupled-channels 
frameworks~\cite{Zhang2013,Zhang2013-2,Kamano2014,Kamano2015,Fernandez2015,Matveev2019,Sarantsev2019,Anisovich2020} 
to perform comprehensive partial-wave analyses of $\bar K N$ reactions with various final states.
Spanning a wide energy range from the threshold up to the $W > 2$~GeV region, 
these studies have successfully achieved a systematic determination of $Y^*$ resonances 
based on the poles of scattering amplitudes in the complex-energy plane.

In this context, we developed a dynamical coupled-channels (DCC) model in 2014 
for meson-baryon reactions in the $S=-1$ sector~\cite{Kamano2014} by applying the theoretical framework of 
the Argonne National Laboratory-The University of Osaka (ANL-Osaka) DCC approach~\cite{Matsuyama2007}.
We determined the model parameters by performing a comprehensive partial-wave analysis of over 17000 data points 
covering unpolarized and polarized observables 
for $K^- p \to MB$ with $MB=\bar K N, \pi\Sigma, \pi\Lambda, \eta\Lambda$, and $K\Xi$.
This procedure yielded two distinct parameter sets, Model~A and Model~B, 
both of which successfully reproduce the existing data from the thresholds up to $W = 2.1$~GeV.
(Hereafter, we refer to this analysis as the 2014 ANL-Osaka DCC analysis or simply the 2014 DCC analysis.)
Subsequently, in Ref.~\cite{Kamano2015}, we searched for $Y^*$ resonance poles within the energy region of 
the partial-wave analysis (from the $\bar K N$ threshold to $W=2.1$~GeV) 
and identified 18 (20) $Y^*$ resonances in Model~A (Model~B).
While we successfully confirmed well-established $Y^*$ resonances in terms of their resonance poles, 
some of the extracted states may correspond to poorly established (one- and/or two-star) resonances 
listed by the Particle Data Group (PDG) or may represent entirely new resonances.
Furthermore, two $J^P=1/2^-$ $\Lambda$ resonances were found just below 
the $\bar K N$ threshold in both Model~A and Model~B, 
which would correspond to the $\Lambda(1380)1/2^-$ and $\Lambda(1405)1/2^-$ states in the PDG~\cite{PDG2024}.

In Refs.~\cite{Kamano2014,Kamano2015}, our spectroscopic study of $Y^*$ resonances was 
primarily focused on the region above the $\bar K N$ threshold.
This focus was inevitable because our DCC model was constructed by analyzing $K^- p$ reaction data; 
thus, a reliable extraction of $Y^*$ resonances is, in principle, limited to the energy region accessible via $K^- p$ reactions.
However, this coverage is insufficient for establishing low-lying $Y^*$ resonances.
Some of these resonances are expected to lie below the $\bar K N$ threshold, 
a region that cannot be directly accessed through $\bar K N$ reactions.
Furthermore, even for low-lying resonances located just above the $\bar K N$ threshold, 
establishing them is difficult because precise measurements of differential cross sections and 
polarization observables for $\bar K N$ reactions 
are very challenging in practice due to the very low momentum of the incoming $\bar K$ beam in this region.

\begin{figure}[tb]
\centering
\includegraphics[clip,width=0.48\textwidth]{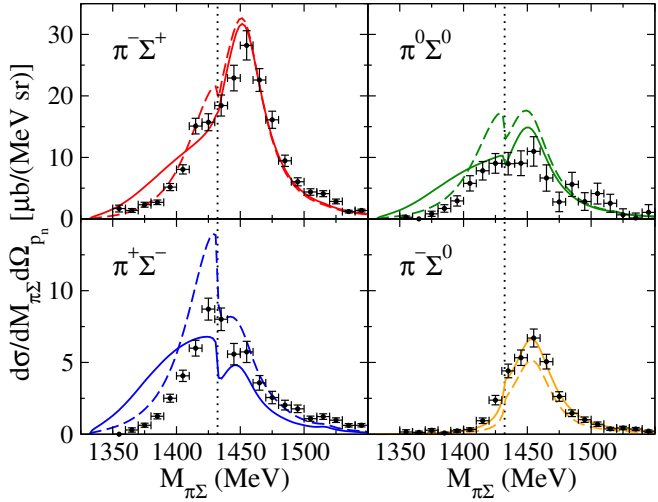}
\caption{
The $\pi\Sigma$ invariant mass distributions for the $K^- d \to \pi\Sigma n$ reaction with the kinematics that the incident $K^-$ momentum is 1~GeV/$c$
and the outgoing neutron is detected at $\theta = 0$.
The solid and dashed curves are the predicted results reported in 2016~\cite{Kamano2016}, where the results are calculated using Model~A and Model~B of 
the ANL-Osaka DCC model of Ref.~\cite{Kamano2014}, respectively.
The data points are from the J-PARC E31 experiment reported in 2023~\cite{J-PARCE31}.
Here, only the total errors are shown for each data point.
The dotted vertical line in each panel represents the $\bar K N$ threshold energy.
}
\label{fig:1000mev-piS}
\end{figure}

To overcome this difficulty, we investigated the $K^- d \to \pi Y N$ reaction in Ref.~\cite{Kamano2016}
as a means to access the energy region below and just above the $\bar{K}N$ threshold.
We calculated the $\pi Y$ invariant mass distributions in this energy region 
by accounting for the impulse and $\bar{K}$-exchange processes (see Fig.~3 in Ref.~\cite{Kamano2016})
and applying Model~A and Model~B of the DCC model~\cite{Kamano2014} to generate 
the off-shell meson-baryon amplitudes that are crucial inputs to the calculations.
These calculations were performed using the same beam energy and kinematics as those of the J-PARC E31 experiment~\cite{J-PARCE31}.
As shown in Fig.~\ref{fig:1000mev-piS}, our predicted results for the $\pi\Sigma$ invariant mass distributions 
reported in 2016~\cite{Kamano2016} reproduce the J-PARC E31 data published in 2023~\cite{J-PARCE31} reasonably well,
particularly in terms of their overall magnitude.
In earlier theoretical studies (see, e.g., Ref.~\cite{Ohnishi2015}; 
the same situation would apply to Refs.~\cite{Jido2009,Jido2012,Yamagata-Sekihara2012,Miyagawa2012}), 
the calculated magnitudes were much smaller than the experimental values.
We found that this discrepancy occurred because these previous studies used inappropriate amplitudes 
for the initial elementary subprocess of the $\bar{K}$-exchange process,
where the incoming $\bar{K}$ meson interacts with the nucleon inside the deuteron
(the $\bar{K} N_2 \to \bar{K}_{\textrm{ex}} N$ subprocess in Fig.~3(b) of Ref.~\cite{Kamano2016}).
We resolved this issue for the first time by utilizing our DCC amplitudes that take into account $S$, $P$, $D$, and $F$ waves, 
while also consistently implementing the off-shell effects of the DCC model.
Moreover, a comparison between the results from Model~A and Model~B revealed that the line shape of 
the $\pi\Sigma$ invariant mass distribution is highly sensitive to the pole positions of 
the two $1/2^-$ $\Lambda$ resonances located just below the $\bar K N$ threshold.
This comparison with the data clearly demonstrates that our appropriate treatment
of the elementary subprocess is indeed critical for explaining the observations.
Consequently, it indicates that the ANL-Osaka DCC model for the $S=-1$ sector is valid for describing observables 
below the $\bar K N$ threshold, and can therefore be reliably used to investigate 
other reactions, such as $\mi \bi \to K^* \mx\bx$, 
in the energy region below and just above the $\bar K N$ threshold.

Given the findings mentioned above, a combined analysis of $\bar K N$ and $\bar K d$ reactions 
is desirable to establish the $Y^*$ mass spectrum.
However, theoretical calculations of $\bar K d$ reactions involving a deuteron target are computationally demanding 
due to the inclusion of nuclear effects and the associated loop integrals.
Consequently, including $\bar K d$ reactions in a combined fit is practically unfeasible.

Therefore, in the present study, we focus on the $\mi\bi \to \mf\mx\bx$ reaction 
as an alternative means to access the energy region below the $\bar K N$ threshold.
Here, the final-state meson-baryon ($\mx\bx$) system is a pair comprising a pseudoscalar meson 
and a spin-$1/2$ baryon with total strangeness $S=-1$.
Importantly, among the possible $\mx\bx$ pairs, the $\pi\Sigma$ and $\pi\Lambda$ channels 
can have an invariant mass lower than the $\bar K N$ threshold energy, 
thereby ensuring that this reaction can effectively probe the subthreshold region.

\begin{figure}[b]
\centering
\includegraphics[clip,width=0.36\textwidth]{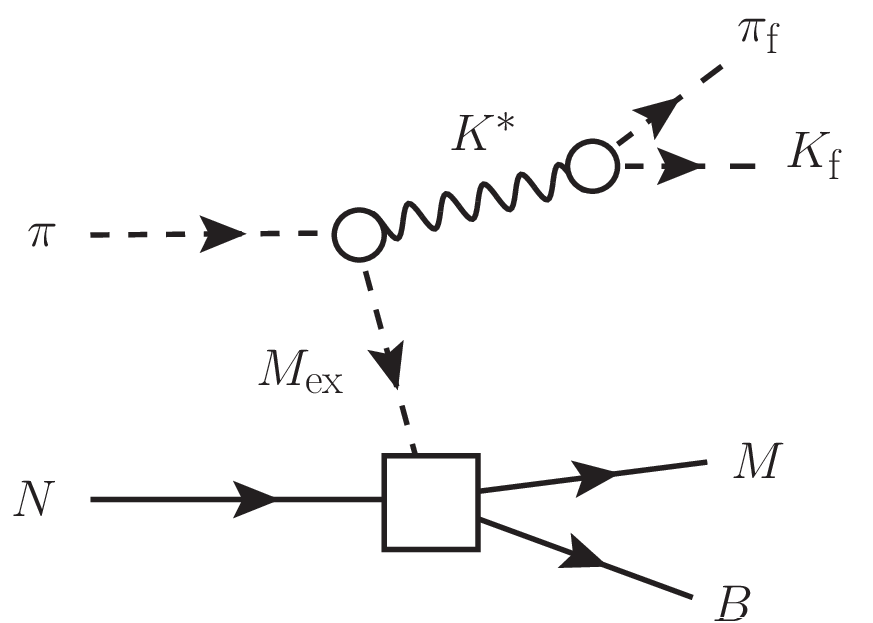}
\caption{
Diagrammatic representation of the $\mi \bi \to \mf \mx \bx$ reaction process considered in this work.
Here, $\mex$ denotes the exchanged meson. 
The instability of the outgoing $\mf$ meson, 
attributed to its strong decay into a $\pi K$ pair (denoted as $\mfo\mft$), 
is properly taken into account.
The half-off-shell amplitudes describing the $\mex \bi \to \mx \bx$ subprocesses 
(indicated by the solid square) are taken from our DCC model of Ref.~\cite{Kamano2014}.
}
\label{fig:reaction1}
\end{figure}

Our strategy is as follows.
We consider kinematics where the incident $\pi$ beam has a high momentum and the outgoing $\mf$ meson is produced at very forward angles.
Under these high-energy peripheral production kinematics, it is well known that the reaction is dominated by 
$t$-channel one-meson exchange processes, as depicted in Fig.~\ref{fig:reaction1}.
The amplitudes for these one-meson exchange processes can be evaluated 
using the half-off-shell amplitudes of the $\mex\bi \to \mx\bx$ subprocesses,
the propagators for the exchanged meson $\mex$, 
and the vertex functions associated with the high momentum $\mi \to \mf\mex$ transitions.
Since the half-off-shell  $\mex \bi \to \mx \bx$ amplitudes can be generated
by solving the coupled-channels scattering equations for $\bar{K}N$ reactions within our DCC approach, 
a combined analysis of the $\bar{K}N$ and $\mi \bi \to \mf\mx\bx$ reactions can be achieved.

Following the strategy outlined above, in this work, we investigate the utility of the $\mi\bi \to \mf\mx\bx$ reaction 
for establishing low-lying $Y^*$ resonances.
Specifically, we examine how the extracted low-lying $Y^*$ states from our DCC analysis (Model~A and Model~B), 
including those located below the $\bar K N$ threshold, manifest in the observables of this reaction.
By comparing the differential cross sections predicted by Model~A and Model~B, 
we identify observables that are sensitive to the existence of the poorly established or newly found resonances within these models.
We note that, among the subthreshold $Y^*$ resonances extracted in our DCC model, 
only the $1/2^-$ $\Lambda$ resonances have been presented in our previous works; 
therefore, we provide a detailed discussion of the other subthreshold resonances in Sec.~\ref{sec:low-ystar}.
The kinematic condition of very forward $\mf$ meson production induced by a high-momentum $\pi$ beam is expected 
to be accessible at hadron facilities such as J-PARC.
We hope that the present results will serve as a motivation for future experimental measurements.

This paper is organized as follows.
In Sec.~\ref{sec:low-ystar}, we present the extracted low-lying $Y^*$ resonances from Model~A and Model~B of 
our DCC analysis~\cite{Kamano2014}, including those lying below the $\bar K N$ threshold.
The kinematics and cross-section formulas used in this work are given in Sec.~\ref{sec:kine-cs}, 
followed by a description of the model for constructing the $\mi\bi \to \mf\mx\bx$ reaction amplitudes in Sec.~\ref{sec:model}.
The results and discussion are presented in Sec.~\ref{sec:results}.
Finally, a summary is given in Sec.~\ref{sec:summary}.

\section{\label{sec:low-ystar}Low-lying $Y^*$ resonances from the 2014 ANL-Osaka DCC analysis}

\begin{table}[b]
\caption{
Low-lying $\Lambda$ and $\Sigma$ resonances extracted from the 2014 DCC analysis of Ref.~\cite{Kamano2014}.
Here, $M_R$ and $J^P$ represent the complex pole mass and spin-parity of the resonances, respectively.
The symbol $l_{I2J}$ is an alternative notation of the resonance quantum numbers, 
given by the orbital angular momentum ($l$), the isospin ($I$), 
and the total angular momentum ($J$) of the corresponding $\bar K N$ partial wave.
The pole masses are listed in the format [$\textrm{Re}(M_R),-\textrm{Im}(M_R)$].
For resonances above the $\bar K N$ threshold, the deduced uncertainties 
for the pole masses evaluated in Ref.~\cite{Kamano2015} are also presented.
}
\label{tab:low-ystar}
\centering
\begin{ruledtabular}
\begin{tabular}{cccccc}
&\multicolumn{2}{c}{Model A} &\multicolumn{2}{c}{Model B} \\
&$M_R$ (MeV) & $J^P(l_{I2J})$ & $M_R$ (MeV) & $J^P(l_{I2J})$ \\
\hline
$\Lambda$ baryons & $(1372,56)$ & $1/2^-(S_{01})$ & $(1397,98)$ & $1/2^-(S_{01})$\\
                  &             &                & $(1401,155)$ & $1/2^+(P_{01})$\\
                  & $(1432,75)$ & $1/2^-(S_{01})$ & $(1428,31)$ & $1/2^-(S_{01})$\\
&&&&\\
$\Sigma$ baryons & $(1303,13)$ & $3/2^+(P_{13})$ & $(1360,38)$ & $3/2^+(P_{13})$\\
                 & $(1381,20)$ & $3/2^+(P_{13})$ & $(1381,20)$ & $3/2^+(P_{13})$\\
                 & $(1381,150)$ & $1/2^-(S_{11})$ & &\\
                 &             &                 & $(1457^{+5}_{-1},39^{+1}_{-4})$ & $1/2^+(P_{11})$\\
                 &             &                 & $(1492^{+4}_{-7},69^{+4}_{-7})$ & $3/2^-(D_{13})$\\
\end{tabular}
\end{ruledtabular}
\end{table}

\begin{figure*}[tb]
\centering
\includegraphics[clip,width=0.8\textwidth]{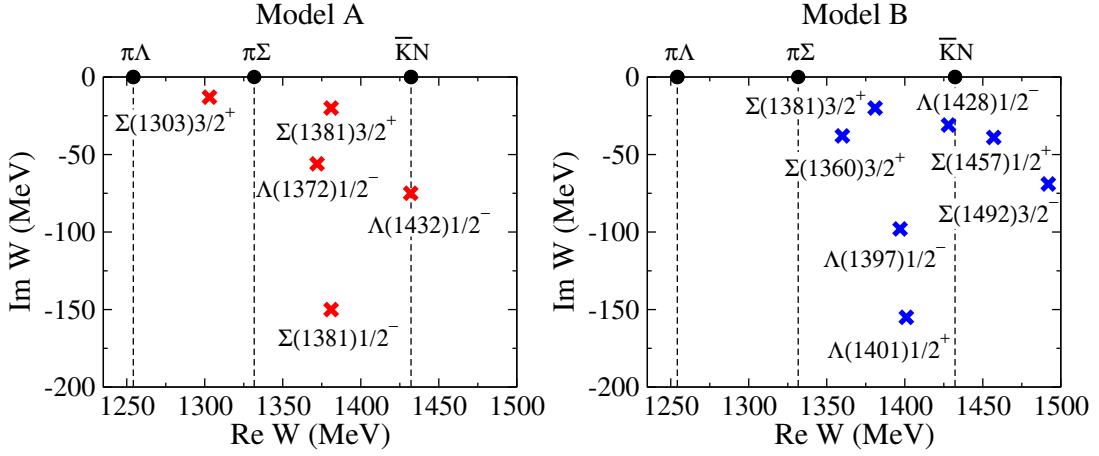}
\caption{
Pole positions of the low-lying $Y^*$ resonances in the complex energy plane.
The left (right) panel shows the results extracted from Model~A (Model~B) of the 2014 DCC analysis~\cite{Kamano2014}.
Crosses represent the resonance pole positions, 
while solid black circles represent the branch points of the $\pi\Lambda$, $\pi\Sigma$, and $\bar{K}N$ channels.
Each cross is labeled with the corresponding resonance name, e.g., $\Sigma(1381)3/2^+$, where the number 
in parentheses denotes the real part of the pole mass and the trailing notation indicates the spin-parity $J^P$.
All presented resonances are located on the sheet nearest to the physical real $W$ axis.
}
\label{fig:low-ystar}
\end{figure*}

\begin{table*}[tb]
\caption{
Residues ($R_{MB,MB}$) of the scattering amplitudes for $MB$ scattering ($MB = \pi\Lambda$ and $\pi\Sigma$) at the resonance pole positions.
The magnitude [$R$ (MeV)] and phase [$\phi$ (degrees), taken to be $-180^\circ < \phi \leq 180^\circ$] of $R_{MB,MB}\equiv R e^{i\phi}$ are listed.
Each resonance is specified by the real part of the pole mass $\mathrm{Re}(M_R)$ and its quantum numbers.
}
\label{tab:residue}
\centering
\begin{ruledtabular}
\begin{tabular}{cccccc}
&&\multicolumn{2}{c}{$R_{\pi\Lambda,\pi\Lambda}$} &\multicolumn{2}{c}{$R_{\pi\Sigma,\pi\Sigma}$} \\
\cline{3-4}\cline{5-6}
&Particle $J^P$($l_{I2J}$)& $R$ & $\phi$ & $R$ & $\phi$\\
\hline
Model A&$\Lambda(1372)$ $1/2^-(S_{01})$&$-$&$-$&$117.97$&$-67$\\ 
       &$\Lambda(1432)$ $1/2^-(S_{01})$&$-$&$-$&$176.80$&$144$\\
       &$\Sigma(1303)$ $3/2^+(P_{13})$ &$13.45$&$-39$&$-$&$-$\\
       &$\Sigma(1381)$ $3/2^+(P_{13})$ &$19.26$&$-107$&$0.54$&$-111$\\
       &$\Sigma(1381)$ $1/2^-(S_{11})$ &$23.00$&$-15$&$62.98$&$179$\\
&&&&&\\
Model B&$\Lambda(1397)$ $1/2^-(S_{01})$&$-$&$-$&$142.38$&$-98$\\ 
       &$\Lambda(1401)$ $1/2^+(P_{01})$&$-$&$-$&$65.99$&$-175$\\
       &$\Lambda(1428)$ $1/2^-(S_{01})$&$-$&$-$&$67.14$&$110$\\
       &$\Sigma(1360)$ $3/2^+(P_{13})$ &$65.81$&$-61$&$1.34$&$26$\\
       &$\Sigma(1381)$ $3/2^+(P_{13})$ &$34.60$&$152$&$3.99$&$-62$\\
       &$\Sigma(1457)$ $1/2^+(P_{11})$ &$40.54$&$-41$&$1.02$&$30$\\
       &$\Sigma(1492)$ $3/2^-(D_{13})$ &$49.70$&$-80$&$6.09$&$-86$\\
\end{tabular}
\end{ruledtabular}
\end{table*}

\begin{table*}[tb]
\caption{
Branching ratios for the decays of low-lying resonances extracted from Model~A and Model~B.
Equations~(23)–(26) in Ref.~\cite{Kamano2015} are used to evaluate the ratios.
See Table~I in Ref.~\cite{Kamano2015} for the quantum numbers of the $(\pi\Sigma^*)_i$ ($i = 1,2$) channels corresponding to a given $J^P$.
}
\label{tab:br-at-real}
\centering
\begin{ruledtabular}
\begin{tabular}{ccccccc}
&&\multicolumn{5}{c}{Branching ratios ($\%$)} \\
\cline{3-7}
&Particle $J^P$($l_{I2J}$)& $B_{\pi \Lambda}$ & $B_{\pi\Sigma}$ & $B_{\bar K N}$ & $B_{(\pi \Sigma^*)_1}$ & $B_{(\pi \Sigma^*)_2}$\\
\hline
Model A&$\Lambda(1372)$ $1/2^-(S_{01})$&$-$&$100.0$&$-$&$-$&$-$\\ 
       &$\Lambda(1432)$ $1/2^-(S_{01})$&$-$&$100.0$&$-$&$0.0$&$-$\\
       &$\Sigma(1303)$ $3/2^+(P_{13})$ &$100.0$&$-$&$-$&$-$&$-$\\
       &$\Sigma(1381)$ $3/2^+(P_{13})$ &$97.6$&$2.4$&$-$&$-$&$-$\\
       &$\Sigma(1381)$ $1/2^-(S_{11})$ &$52.3$&$47.7$&$-$&$-$&$-$\\
&&&&&&\\
Model B&$\Lambda(1397)$ $1/2^-(S_{01})$&$-$&$100.0$&$-$&$0.0$&$-$\\ 
       &$\Lambda(1401)$ $1/2^+(P_{01})$&$-$&$100.0$&$-$&$0.0$&$-$\\
       &$\Lambda(1428)$ $1/2^-(S_{01})$&$-$&$100.0$&$-$&$0.0$&$-$\\
       &$\Sigma(1360)$ $3/2^+(P_{13})$ &$96.9$&$3.1$&$-$&$-$&$-$\\
       &$\Sigma(1381)$ $3/2^+(P_{13})$ &$87.9$&$12.1$&$-$&$-$&$-$\\
       &$\Sigma(1457)$ $1/2^+(P_{11})$ &$96.7$&$2.1$&$1.2$&$0.0$&$-$\\
       &$\Sigma(1492)$ $3/2^-(D_{13})$ &$90.7$&$9.2$&$0.0$&$0.1$&$0.0$\\
\end{tabular}
\end{ruledtabular}
\end{table*}

In this section, we present the low-lying $Y^*$ resonances 
extracted from Model~A and Model~B of our 2014 DCC analysis~\cite{Kamano2014}.
In this paper, the term ``low-lying $Y^*$ resonances'' refers to 
those states whose pole masses have real parts lower than $1.5$~GeV.
This low-lying $Y^*$ resonance region corresponds to
energies below and just above the $\bar K N$ threshold,
where the $\mi \bi \to \mf \mx \bx$ reaction is expected to play a key role in establishing
these states, as discussed in Sec.~\ref{sec:intro}.

Of course, our 2014 DCC analysis also identified poorly established and/or new resonances 
in the energy region above $1.5$~GeV.
However, for such high-lying $Y^*$ resonances, one can, 
in principle, utilize high-statistics data from so-called complete experiments on $\bar K N \to M(0^-)B(1/2^+)$, 
in which angular distributions and all possible polarizations
(the polarization $P$ and the spin rotations) are measured, 
as well as differential cross sections for two-meson production channels.
Thus, those conventional $\bar K N$ reactions should provide the best avenues for establishing these higher-mass states.

In Table~\ref{tab:low-ystar}, we present the complex pole masses 
of the low-lying $Y^*$ resonances extracted from Model~A and Model~B of our 2014 DCC analysis~\cite{Kamano2014}.
The corresponding positions of the resonance poles in the complex energy plane are shown in Fig.~\ref{fig:low-ystar}.
In the low-lying $Y^*$ resonance region, we have identified a total of 5 and 7 $Y^*$ resonances in Model~A and Model~B, respectively.
Note that some of these states have already been reported in Ref.~\cite{Kamano2015}: 
specifically, the two $1/2^-$ $\Lambda$ resonances found in both Model~A and Model~B 
(which would correspond to the $\Lambda(1380)1/2^-$ and $\Lambda(1405)1/2^-$ in the PDG~\cite{PDG2024}), 
as well as the $1/2^+$ and $3/2^-$ $\Sigma$ resonances found just above the $\bar K N$ threshold in Model~B.
Regarding the latter $1/2^+$ and $3/2^-$ $\Sigma$ resonances in Model~B, their counterparts are also found in Model~A; 
however, because the real parts of their pole masses in Model~A are higher than $1.5$~GeV, they are not presented here.
These resonances are assigned to the three-star $\Sigma(1660)1/2^+$ and the one-star $\Sigma(1580)3/2^-$ in the PDG~\cite{PDG2024}, respectively.
On the other hand, the remaining resonances listed in the table are presented here for the first time.

A striking feature of our results is that both Model~A and Model~B exhibit two $3/2^+$ $\Sigma$ resonances in the energy region below the $\bar K N$ threshold.
One of them has a pole mass of $1381-i20$~MeV, corresponding to the well-established $\Sigma(1385)3/2^+$ in the PDG~\cite{PDG2024}.\footnote{
It should be noted that because our 2014 analysis solely utilized $\bar K N$ reaction data, 
the models were essentially constrained only in the energy region above the $\bar K N$ threshold.
However, recognizing the firmly established existence of the $\Sigma(1385)3/2^+$, 
we explicitly imposed its pole position ($1381-i20$)~MeV, obtained by averaging the $\Sigma^{*+}$ and 
$\Sigma^{*-}$ pole masses from Refs.~\cite{Lichtenberg1974,PDG2024} as an additional constraint in the $\chi^2$ fitting of our 2014 analysis.}
The other $3/2^+$ resonance pole, located at $1303-i13$~MeV ($1360-i38$~MeV) in Model~A (Model~B), represents a completely new state.
In Model~A, the pole of this new $3/2^+$ $\Sigma$ resonance lies between the $\pi\Lambda$ and $\pi\Sigma$ thresholds, 
and its real part is approximately $80$~MeV lower than that of the $\Sigma(1385)3/2^+$.
Furthermore, it is significantly narrower than the $\Sigma(1385)3/2^+$.
On the other hand, in Model~B, the pole of this new state is located much closer to the $\Sigma(1385)3/2^+$.
We will investigate how these differences in the properties of the new state between Model~A and Model~B 
manifest in the cross sections of the $\mi \bi \to \mf\mx\bx$ reaction in Sec.~\ref{sec:results}.

In addition to the new $3/2^+$ $\Sigma$ resonance, we have identified other potentially new $Y^*$ resonances 
in the energy region below the $\bar K N$ threshold: 
a $1/2^-$ $\Sigma$ resonance with a pole mass of $1381-i150$~MeV in Model~A, and 
a $1/2^+$ $\Lambda$ resonance with a pole mass of $1401-i155$~MeV in Model~B.
Although these states have large imaginary parts compared to other low-lying $Y^*$ resonances and are not common to both models, 
investigating how their presence or absence affects the cross sections of the $\mi\bi \to \mf \mx\bx$ reactions may yield valuable insights.
In particular, the possible existence of a low-lying $1/2^-$ $\Sigma$ resonance has been widely investigated 
both experimentally and theoretically (see, e.g., Ref.~\cite{Wang2024,Belle2022,He2025}),
and the $\mi\bi \to \mf \mx\bx$ reaction data could provide a further opportunity to clarify the nature of this controversial resonance.

In Tables~\ref{tab:residue} and~\ref{tab:br-at-real}, 
we present the residues and branching ratios of low-lying $Y^*$ resonances, respectively.
For the details of the definition of the residues and branching ratios, see Ref.~\cite{Kamano2015}.

\section{\label{sec:kine-cs}Kinematics and cross sections}

In this section, we present kinematics and the cross section formulas used in this work.
Denoting the incoming and outgoing hadrons as shown in Fig.~\ref{fig:reaction1},
the cross section for the $\mi \bi \to \mfo \mft \mx \bx$ reaction is given by
\begin{align}
d\sigma = 
&
 \frac{1}{4\sqrt{(p_\mi\cdot p_\bi)^2 -\mmi^2 \mbi^2}}
 \frac{d \moms{p}_{\mfo}}{{\cal N}_{\mfo}}
 \frac{d \moms{p}_{\mft}}{{\cal N}_{\mft}}
 \frac{d \moms{p}_{\mx}}{{\cal N}_{\mx}}
 \frac{d \moms{p}_{\bx}}{{\cal N}_{\bx}}
\nonumber\\
&\times
 (2\pi)^4
 \delta^4 \left(p_\mi + p_\bi - p_{\mfo} - p_{\mft} - p_\mx - p_\bx \right) 
\nonumber\\
&\times
 \left|{\cal M}_{\mfo\mft MB,\mi\bi}\right|^2 ,
\label{eq:crs-general}
\end{align}
where $m_i$, $p_i$, and $\moms{p}_i$ denote the mass, four-momentum, and three-momentum of hadron $i$, respectively;
the factor ${\cal N}_i$ is given by ${\cal N}_i = (2\pi)^3 2 E_i(\moms{p}_i)$ 
with $E_i(\moms{p}_i)$ being the relativistic energy of hadron $i$;
and ${\cal M}_{\mfo\mft MB,\mi\bi}$ is the invariant amplitude for
the $\mi \bi \to \mfo \mft \mx \bx $ reaction.
Here, we adopt the Lorentz-invariant normalization for one-particle states,
$\inp{\moms{p}'}{\moms{p}}=(2\pi)^3 2E(\moms{p})\delta^3(\moms{p} -\moms{p}')$,
and the convention that the Dirac spinor $u$ is normalized as $\bar u u = 2 m$.

As mentioned in Sec.~\ref{sec:intro}, in this work 
we consider the case where the $\mi \bi \to \mfo \mft \mx \bx$ reaction 
proceeds via the following cascade process, 
in which the final $\mfo$ and $\mft$ mesons are produced via the decays of $\mf$:
\begin{align}
\mi \bi \to \mf \mx \bx; \quad \mf \to \mfo \mft.
\end{align}
In this case, the invariant amplitude for the reaction can be decomposed as
\begin{align}
i{\cal M}_{\mfo \mft \mx \bx, \mi \bi} = 
&
 \sum_{\lambda_\mf} 
 i{\cal M}_{\mfo\mft,\mf(\lambda_\mf)}
\times 
 iD_\mf
\nonumber\\
&\times 
 i{\cal M}_{\mf(\lambda_\mf) \mx\bx, \mi \bi},
\label{eq:model-amp}
\end{align}
where $\lambda_\mf$ denotes the helicity of $\mf$; and
$i{\cal M}_{\mfo\mft,\mf(\lambda_\mf)}$, $iD_\mf$, and $i{\cal M}_{\mf(\lambda_\mf) \mx\bx, \mi \bi}$
represent the invariant amplitude for $\mf\to \mfo \mft$, 
the propagator of $\mf$, 
and the invariant amplitude for $\mi \bi\to\mf \mx \bx$, respectively.
Note that since these factors are individually invariant under the Lorentz boost,
each of them can be evaluated in a different Lorentz frame.
In this work, we evaluate $i{\cal M}_{\mfo\mft,\mf(\lambda_\mf)}$ and $iD_\mf$ in the rest frame of $\mf$.

In the rest frame of $\mf$, ${\cal M}_{\mfo\mft, \mf(\lambda_\mf)}$ can be factorized as
\begin{align}
{\cal M}_{\mfo\mft, \mf(\lambda_\mf)} =
&
 \cg{I_\mfo I^z_\mfo}{I_\mft I^z_\mft}{I_\mf I^z_\mf}
\nonumber\\
&\times
 Y_{1\lambda_\mf}(\moms{\hat\kappa})
 \tilde{\cal M}_{\mfo\mft,\mf}(|\moms{\kappa}|),
\label{eq:vp1p2-vtx}
\end{align}
where $I_i$ and $I^z_i$ denote the isospin and its projection of hadron $i$ ($i=\mfo, \mft, \mf$), respectively.
The three-momentum of $\mfo$ in the rest frame of $\mf$ is denoted by $\moms{\kappa}$ 
(with its direction $\moms{\hat \kappa}$),
whose magnitude is given by $|\moms{\kappa}|=[\lambda(\mmf^2,\mmfo^2,\mmft^2)]^{1/2}/(2\mmf)$.
Here, $\mmf$ is the invariant mass of the $\mfo\mft$ system, and $\lambda(a,b,c)$ is the K\"{a}ll\'en function defined by
$\lambda(a,b,c) = a^2 + b^2 + c^2 -2ab -2bc -2ca$.

Substituting Eqs.~(\ref{eq:model-amp}) and~(\ref{eq:vp1p2-vtx}) into Eq.~(\ref{eq:crs-general}), 
and performing some manipulation,
we obtain the following expressions for the unpolarized differential cross section:
\begin{align}
\frac{d\sigma}{dtdWd\Omega^*_\mx} =
&
 \cg{I_\mfo I^z_\mfo}{I_\mft I^z_\mft}{I_\mf I^z_\mf}^2
\nonumber\\
&\times
 \int_{M_{\mf}^{\rm{min}}}^{M_{\mf}^{\rm{max}}} d\mmf 
 {\cal W}(\mmf)
 \frac{d\sigma_{\mf \mx\bx,\mi\bi}}{dtdWd\Omega^*_\mx},
\label{eq:crs-with-iso}
\end{align}
with
\begin{align}
\frac{d\sigma_{\mf \mx\bx,\mi\bi}} {dt dW d\Omega^*_\mx} =
&
 \frac{|\moms{p}_\mx^*|}{(4\pi)^4 \lambda(s,\mmi^2,\mbi^2)}
 \overline{\sum_{\textrm{spins}}} |{\cal M}_{\mf\mx\bx,\mi \bi}|^2 ,
\label{eq:crs-pip-ksmb}
\end{align}
and
\begin{align}
{\cal W}(\mmf) =
&
  \frac{|\moms{\kappa}|}{16\pi^3} 
  |\tilde{\cal M}_{\mfo\mft,\mf}(|\moms{\kappa}|)|^2 |D_\mf|^2 .
\label{eq:weight}
\end{align}
Here, $t$ is the squared momentum transfer $q$, defined as $t=q^2$ with $q = p_\mi - p_\mf$ and $p_\mf = p_\mfo + p_\mft$;
$W$ is the invariant mass of the $\mx\bx$ system, given by $W=\sqrt{(p_\mx + p_\bx)^2}$;
$\moms{p}_\mx^*$ is the three-momentum of $\mx$ in the $\mx\bx$ center-of-mass frame,
whose magnitude is given by $|\moms{p}_M^*| = [\lambda(W^2,\mmx^2,\mbx^2)]^{1/2}/(2W)$;
and $\Omega^*_\mx$ is the solid angle of $\moms{p}_M^*$.
(Hereafter, the three-momenta and other kinematical variables marked with an asterisk denote those in the center-of-mass frame of the $\mx\bx$ system.)
In this work, the coordinate system in the $\mx\bx$ center-of-mass frame is defined
such that the $z$-axis is parallel to $\moms{q}^*$, and the $x$-$z$ plane is spanned by the momenta of the incident pion beam and the outgoing $\mf$.
The symbol $\overline{\sum}_{\textrm{spins}}$ in Eq.~(\ref{eq:crs-pip-ksmb}) 
denotes the summation over the spins of the final-state particles ($\mf$, $\mx$, and $\bx$) and the average over the initial nucleon spin.

Equation~(\ref{eq:crs-with-iso}) shows that, apart from the isospin factor, the cross section consists of two parts:
$d\sigma_{\mf \mx\bx,\mi\bi}/(dt dW d\Omega^*_\mx)$ [Eq.~(\ref{eq:crs-pip-ksmb})] and ${\cal W}(\mmf)$ [Eq.~(\ref{eq:weight})].
The former corresponds to the unpolarized cross section for the reaction $\mi N \to \mf MB$ treated as if $\mf$ were a stable particle against the strong interaction.
Consequently, ${\cal W}(\mmf)$ acts as a mass-distribution (weight) function.
The factor ${\cal W}(\mmf)$ together with the integration over $\mmf$ in Eq.~(\ref{eq:crs-with-iso}) 
accounts for the finite width of $\mf$ due to the $\mf \to \mfo\mft$ decay.

\begin{figure}[t]
\centering
\includegraphics[clip,width=0.3\textwidth]{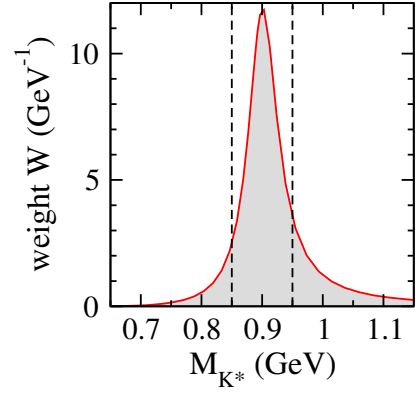}
\caption{
The weight function ${\cal W}(M_\mf)$ obtained with the $I=1/2$ and $L=1$ $\pi K$ scattering model described in Appendix~\ref{app:weight}.
The range between the two vertical dashed lines ($0.85\leq M_\mf \leq 0.95$~GeV) 
is considered for the $M_\mf$ integration in computing the cross sections for $\mi \bi \to \mf \mx\bx$
[Eq.~(\ref{eq:crs-dtdwdo})].
}
\label{fig:weight}
\end{figure}

Since $\mf$ decays into $\pi K$ with a branching fraction of almost $100 \%$~\cite{PDG2024},
we define the cross section for the $\pi N \to K^{*+} M B$ process as the sum of the 
$\pi N \to (\pi^+ K^0) M B$ and $\pi N \to (\pi^0 K^+) M B$ channels.
Similarly, the $\pi N \to K^{*0} M B$ process is identified as the sum of the 
$\pi N \to (\pi^- K^+) M B$ and $\pi N \to (\pi^0 K^0) M B$ channels.
Based on these definitions and using 
$\sum_{I^z_\mfo,I^z_\mft} \cg{1 I^z_\mfo}{\frac{1}{2} I^z_\mft}{\frac{1}{2} +\frac{1}{2}}^2 = 1$ and
$\sum_{I^z_\mfo,I^z_\mft} \cg{1 I^z_\mfo}{\frac{1}{2} I^z_\mft}{\frac{1}{2} -\frac{1}{2}}^2 = 1$,
we finally obtain the unpolarized differential cross section 
for the $\pi N \to K^* M B$ reaction, in which the finite width effects of the outgoing $K^*$ are properly incorporated:
\begin{align}
\frac{d\sigma}{dtdWd\Omega^*_\mx} =
&
 \int_{M_{\mf}^{\rm{min}}}^{M_{\mf}^{\rm{max}}} d\mmf 
 {\cal W}(\mmf)
 \frac{d\sigma_{\mf \mx\bx,\mi\bi}}{dtdWd\Omega^*_\mx}.
\label{eq:crs-dtdwdo}
\end{align}
In principle, the integration limits in Eq.~(\ref{eq:crs-dtdwdo}) are ${M_{\mf}^{\mathrm{min}}} = \mmfo + \mmft$ and $M_{\mf}^{\mathrm{max}}=\infty$.
In practice, however, these limits are chosen to cover the range where the resonance peak of $\mf$ is clearly observed in ${\cal W}(M_\mf)$.
A detailed description of the model used to evaluate ${\cal W}(M_\mf)$ is given in Appendix~\ref{app:weight}.
As shown in Fig.~\ref{fig:weight}, the weight function calculated with this model exhibits a clear resonance peak within the range of $0.85$--$0.95$~GeV.
Accordingly, we set the integration limits to ${M_{\mf}^{\mathrm{min}}} = 0.85$~GeV and $M_{\mf}^{\mathrm{max}}=0.95$~GeV.
The calculation of $d\sigma/(dtdW)$ and $d\sigma/dt$ from Eq.~(\ref{eq:crs-dtdwdo}) is straightforward.

\section{\label{sec:model}Model for the $\pi N \to K^* M B$ reaction}

In this section, we describe the model used to evaluate the invariant amplitude 
${\cal M}_{\mf\mx\bx,\mi \bi}$ in Eq.~(\ref{eq:crs-pip-ksmb}).
As mentioned in Sec.~\ref{sec:intro}, in this work 
we assume that the reaction proceeds exclusively via 
the one-meson exchange processes as depicted in Fig.~\ref{fig:reaction1}.
It is known that this assumption is well justified for the very forward $K^*$ production kinematics 
considered in this study (see, e.g., Refs.~\cite{Kim2015,Xiang2020}).
We consider two types of one-meson exchange processes involving $\bar K(0^-)$ and $\bar K^*(1^-)$ mesons.
(These processes are referred to as the $\pex$-exchange process and the $\vex$-exchange process, respectively.)
The invariant amplitude for $\mi \bi \to \mf \mx \bx$ is then expressed as
\begin{align}
i{\cal M}_{\mf\mx\bx,\mi\bi} =
& 
 \sum_{\mex=\pex,\vex}i{\cal M}^{(\mex)}_{\mf\mx\bx,\mi\bi}.
\end{align}
Here, ${\cal M}^{(\mex)}_{\mf\mx\bx,\mi\bi}$ denotes the invariant amplitude for the $\mex$-exchange process.
This amplitude can be decomposed into three factors:
\begin{align}
i{\cal M}_{\mf \mx\bx,\mi\bi}^{(\mex)} =
&
 \sum_{\mex~\mathrm{d.o.f.}}
 i{\cal M}_{\mx\bx, \mex \bi} 
 \times 
 iD_\mex
\nonumber\\
&\times 
 i{\cal M}_{\mf\mex, \mi},
\label{eq:mexamp_general_form}
\end{align}
where ${\cal M}_{\mx\bx, \mex \bi}$, $D_\mex$, and ${\cal M}_{\mf \mex, \mi}$ represent 
the invariant amplitude for $\mex \bi \to \mx \bx$,
the propagator of $\mex$,
and the invariant amplitude for $\mi \to \mf \mex$, respectively; and
the summation runs over the internal degrees of freedom (spin and isospin projections) of the exchanged meson $\mex$.
Since these factors are individually invariant under Lorentz boosts, 
we evaluate ${\cal M}_{\mx\bx, \mex \bi}$ in the center-of-mass frame of the $\mx\bx$ system, while
${\cal M}_{\mf \mex, \mi}$ and $D_\mex$ are evaluated in the laboratory frame.

According to Eq.~(\ref{eq:mexamp_general_form}), the invariant amplitudes for the $\pex$-exchange and $\vex$-exchange processes are
explicitly given by
\begin{widetext}
\begin{align}
&
i{\cal M}_{\mf(\lambda_\mf, I^z_{\mf}) \mx(I^z_M)\bx(S^z_\bx,I^z_\bx),\mi(I^z_\mi) \bi(S^z_\bi,I^z_\bi)}^{(\pex)} (p_\mf,p_\mx,p_\bx ;p_\mi,p_\bi) =
\nonumber\\
&\quad
 \sum_{I_{\pex}^z}
 i{\cal M}_{\mx(I^z_\mx)\bx(S^z_\bx,I^z_\bx), \pex(I^z_\pex) \bi(S^z_\bi,I^z_\bi)}(p_\mx,p_\bx;p_\pex,p_\bi)
\times 
 iD_\pex
\times 
 i{\cal M}_{\mf(\lambda_\mf,I^z_\mf) \pex(I^z_\pex), \mi(I^z_\mi)}(p_\mf,p_\pex;p_\mi) ,
\label{eq:amp-pex}
\end{align}
and
\begin{align}
&
i{\cal M}_{\mf(\lambda_\mf,I^z_\mf) \mx(I^z_\mx)\bx(S^z_\bx,I^z_\bx),\mi(I^z_\mi) \bi(S^z_\bi,I^z_\bi)}^{(\vex)} (p_\mf,p_\mx,p_\bx ;p_\mi,p_\bi) =
\nonumber\\
&\quad
 \sum_{S_{\vex}^z,I_{\vex}^z}
 i{\cal M}_{\mx(I^z_\mx)\bx(S^z_\bx,I^z_\bx), \vex(S^z_{\vex},I^z_\vex) \bi(S^z_\bi,I^z_\bi)}(p_\mx,p_\bx;p_\vex,p_\bi)
 \times 
 iD_\vex
\nonumber\\
&\quad\qquad\qquad
 \times 
 i{\cal M}_{\mf(\lambda_\mf,I^z_\mf) \vex(S^z_\vex,I^z_\vex), \mi(I^z_\mi)}(p_\mf,p_\vex;p_\mi) .
\label{eq:amp-vex}
\end{align}
\end{widetext}
Here, $S^z_X$ ($I^z_X$) and $p_X$ denote the spin (isospin) projection and four-momentum of the particle $X$, respectively.

In the center-of-mass frame of the $\mx\bx$ system, the partial-wave expansion of $i{\cal M}_{\mx\bx, \mex \bi}$
is given for the $\mex =\pex$ case by
\begin{widetext}
\begin{align}
& 
i{\cal M}_{\mx(I^z_\mx)\bx(S^z_\bx,I^z_\bx), \pex(I^z_\pex) \bi(S^z_\bi,I^z_\bi)}(p_\mx,p_\bx;p_\pex,p_\bi) = 
\nonumber\\
&\quad
\sum_{I}
\cg{I_\mx I_\mx^z} {I_\bx I_\bx^z}{I I_\mx^z+I_\bx^z}
\cg{I_\pex I_\mx^z+I_\bx^z - I_\bi^z} {I_\bi I_\bi^z}{I I_\mx^z+I_\bx^z}
\nonumber\\
&\qquad
\times
\sum_{LL^{z}L'^{z}JJ^z}
\cg{LL'^{z}}{S_\bx S^z_\bx}{JJ^z}
Y_{LL'^{z}}(\moms{\hat p}_\mx^*)
\cg{LL^z}{S_\bi S^z_\bi}{JJ^z}
Y_{LL^{z}}^*(\moms{\hat p}_\pex^*)
\times 
i\tilde{\cal M}^{(ILJ)}_{\mx\bx, \pex \bi}(|\moms{p}_\mx^*|,|\moms{p}_\pex^*|;W),
\label{eq:amp-pexNmb}
\end{align}
and for the $\mex =\vex$ case by
\begin{align}
&
i{\cal M}_{\mx(I^z_\mx)\bx(S^z_\bx,I^z_\bx), \vex(S^z_\vex,I^z_\vex) \bi(S^z_\bi,I^z_\bi)}(p_\mx,p_\bx;p_\vex,p_\bi) =
\nonumber\\
&\qquad
\sum_{I}
\cg{I_\mx I_\mx^z} {I_\bx I_\bx^z}{I I_\mx^z+I_\bx^z}
\cg{I_\vex I_\mx^z+I_\bx^z - I_\bi^z} {I_\bi I_\bi^z}{I I_\mx^z+I_\bx^z}
\nonumber\\
&\quad\qquad\qquad
\times
\sum_{L'L'^{z}SS^zLL^{z}JJ^z}
\cg{L'L'^{z}}{S_\bx S^z_\bx}{JJ^z}
Y_{L',L'^{z}}(\moms{\hat p}_\mx^*)
\cg{S_\vex S^z_\vex}{S_\bi S^z_\bi}{SS^z}
\cg{LL^z}{S S^z}{JJ^z}
Y_{L,L^{z}}^*(\moms{\hat p}_\vex^*)
\nonumber\\
&\quad\qquad\qquad
\times
i\tilde{\cal M}^{(IL'LSJ)}_{\mx\bx, \vex \bi}(|\moms{p}_\mx^*|,|\moms{p}_\vex^*|;W).
\label{eq:amp-vexNmb}
\end{align}
\end{widetext}
For the half-off-shell partial-wave amplitudes
of the $\pex N \to \mx \bx$ and $\vex N \to \mx \bx$ processes 
[$\tilde{\cal M}^{(ILJ)}_{\mx\bx, \pex \bi}(|\moms{p}_\mx^*|,|\moms{p}_\pex^*|;W)$ and
$\tilde{\cal M}^{(IL'LSJ)}_{\mx\bx, \vex \bi}(|\moms{p}_\mx^*|,|\moms{p}_\vex^*|;W)$],
we employ those generated from the DCC model of the meson-baryon reactions in the strangeness $S=-1$ sector
developed in Ref.~\cite{Kamano2014}.
It is worth noting that this model consistently accounts for off-shell effects
by solving the full scattering equation without introducing any on-shell approximation.
Specifically, the partial-wave amplitudes, denoted as $\tilde{\cal M}_{\mx\bx,\mex\bi}$ for simplicity, are related to those
obtained from our DCC model ($T^{\textrm{DCC}}_{\mx\bx, \mex \bi}$), as follows:
\begin{align}
i\tilde{\cal M}_{\mx\bx, \mex \bi} =
&
-\frac{i}{(2\pi)^3}
\nonumber\\
&\times
\sqrt{(2\pi)^3 2E_\mx(\moms{p}_\mx^*)}
\sqrt{(2\pi)^3 2E_\bx(\moms{p}_\bx^*)}
\nonumber\\
&\times
\sqrt{(2\pi)^3 2E_\mex (\moms{p}_\mex^*)}
\sqrt{(2\pi)^3 2E_\bi (\moms{p}_\bi^*)}
\nonumber\\
&\times
T^{\textrm{DCC}}_{\mx\bx, \mex \bi}.
\end{align}

For the propagator of the exchanged $\mex$ meson, $iD_\mex$, 
we employ the standard Feynman propagator given by
\begin{align}
iD_\mex = 
&
 \frac{i}{p_\mex^2 - \mmex^2}.
\end{align}
In peripheral high-energy hadron production reactions, such as the one considered here,
a Reggeized propagator may be employed for $iD_\mex$.
The differences between applying Feynman and Reggeized propagators for exchanged mesons
have been discussed in the literature (see, e.g., Refs.~\cite{Kim2015,Xiang2020}).
However, as demonstrated in Refs.~\cite{Kim2015,Xiang2020}, 
regardless of the choice of propagator, reproducing existing data requires 
fine-tuning of additional parameters introduced in the form factors for the $\mi \to \mf \mex$ vertices
(or parameters included in the Reggeized propagator).
These phenomenological parts containing additional parameters are independent of the DCC model.
In our framework, these parameters should ultimately be determined by fitting to future experimental data.
Since such a detailed analysis is beyond the scope of this paper, 
we adopt the standard Feynman propagator in this work.

The invariant amplitude for the $\mi \to \mf \mex$ vertex is given 
for the $\mex = \pex$ case by
\begin{align}
&
i{\cal M}_{\mf(\lambda_\mf,I^z_\mf) \pex(I^z_\pex), \mi(I^z_\mi)}(p_\mf,p_\pex;p_\mi) =
\nonumber\\
&\qquad
 i \cg{I_\mf I^z_{\mf}}{I_\pex I^z_\pex}{I_\mi I^z_\mi} 
 \left(-\frac{1}{\sqrt{2}}\right)\left(-\frac{g''}{\sqrt{24}}\right)
\nonumber\\
&\qquad
 \times
 (p_\mi+p_\pex)_\mu [\varepsilon^{(\lambda_\mf)\mu}_{\mf}]^*  ,
\label{eq:mimfpex-vtx}
\end{align}
and for the $\mex = \vex$ case by
\begin{align}
&
i{\cal M}_{\mf(\lambda_\mf,I^z_\mf) \vex(S^z_\vex,I^z_\vex), \mi(I^z_\mi)}(p_\mf,p_\vex;p_\mi) =
\nonumber\\
&\qquad
 i \cg{I_\mf I^z_{\mf}}{I_\vex I^z_\vex}{I_\mi I^z_\mi}
 \left(-\frac{3}{\sqrt{2}}\right)\left(-\frac{g'''}{\sqrt{30}}\right)
\nonumber\\
&\qquad
\times
 \epsilon^{\alpha\beta\gamma\delta} 
 (+ip_{\mf\alpha})[\varepsilon^{(\lambda_\mf)}_{\mf\beta}]^*
 (+ip_{\vex\gamma})[\varepsilon^{(S^z_{\vex})}_{\vex\delta}]^* ,
\label{eq:mimfvex-vtx}
\end{align}
where $\varepsilon^{(\lambda_\mf)\mu}_{\mf}$ and $\varepsilon^{(S^z_{\vex})}_{\vex\delta}$ denote
the polarization vectors for the outgoing $\mf$ meson and the exchanged $\vex$ meson, respectively.
These expressions are derived from the phenomenological SU(3) Lagrangians (see Appendix~\ref{sec:vpp-vvp-vtx} for details).
In this work, the coupling constant for the $\mi \to \mf \pex$ vertex is taken from Ref.~\cite{Oh2004} 
to be $-g''/\sqrt{24} = 6.56$, which corresponds to $g_{K^* K\pi}$ in Ref.~\cite{Oh2004}.
For the coupling constant of the $\mi \to \mf \vex$ vertex, $g'''$, 
we rely on the SU(3) relation where $-g'''/\sqrt{30}$ corresponds to $g_{\rho\omega\pi}=14.9$~GeV$^{-1}$ in Ref.~\cite{Oh2004};
we adopt this value in the present work.
To account for the finite size of hadrons,
we introduce a phenomenological form factor of the form $(\Lambda_{\mex}^2 - \mmex^2)/(\Lambda_{\mex}^2 - t)$
for both the $\mex = \pex$ and $\mex = \vex$ cases,
and multiply the vertex amplitudes by this factor.
The cutoff parameters $\Lambda_{\pex}$ and $\Lambda_{\vex}$ will be discussed in Sec.~\ref{sec:results}.

\section{\label{sec:results}Results and discussion}

\begin{figure}[tb]
\centering
\includegraphics[clip,width=0.48\textwidth]{pimp2KsSs-395.eps}
\caption{
Differential cross section $d\sigma/dt$ for $\pi^- p \to K^{*0} \Sigma^{*0}(1385)$ at $|\moms{p}_\mi^\lfr|=3.95$~GeV/$c$.
The data are from Ref.~\cite{CERN-CFMS1980}.
Left (right): Results calculated using Model~A (Model~B) of Ref.~\cite{Kamano2014} 
for evaluating the $\mex \bi \to \mx \bx$ subprocesses.
The red solid curves represent the full results, while the dashed (dot-dashed) curves denote the results
where only the $\pex$-exchange ($\vex$-exchange) process is taken into account.
}
\label{fig:pimp2KsSs}
\end{figure}
\begin{figure*}[tb]
\centering
\includegraphics[clip,width=0.8\textwidth]{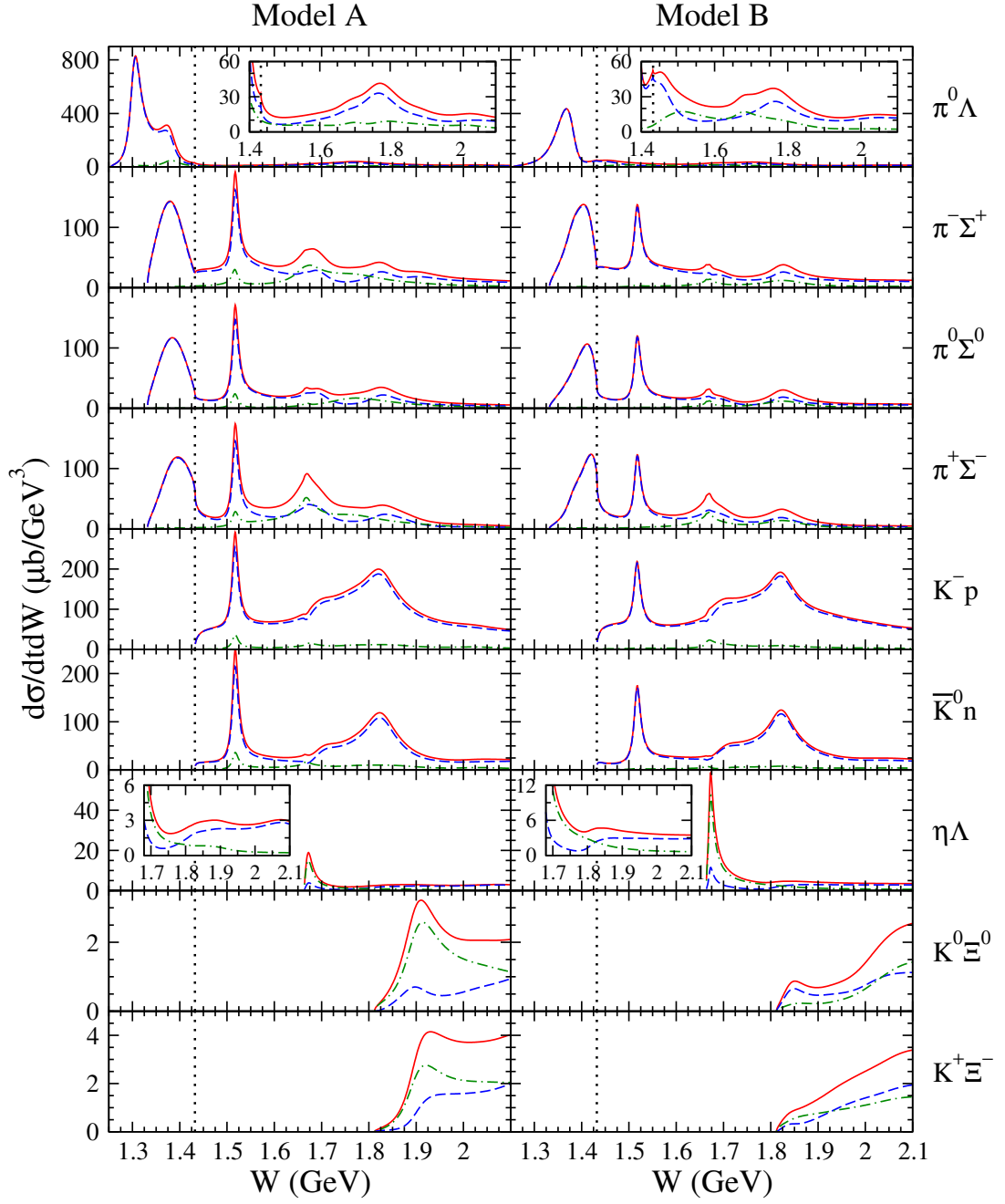}
\caption{
Differential cross section $d\sigma/(dtdW)$ for the $\pi^- p \to K^{*0} \mx \bx$ reactions at $|\moms{p}_\mi^\lfr| = 10$~GeV/$c$ and $t'=0$.
The left (right) panels show the results from Model~A (Model~B).
Each row corresponds to a specific $\mx\bx$ state, as indicated on the right side of the row.
The insets in the $\mx\bx=\pi^0\Lambda$ ($\mx\bx=\eta\Lambda$) panels show an enlarged view of the $W\geq 1.4$~GeV ($W\geq 1.68$~GeV) region.
The meaning of each curve is the same as in Fig.~\ref{fig:pimp2KsSs}.
The dotted vertical line in each panel represents the $\bar K N$ threshold energy.
}
\label{fig:dsdtdw-all}
\end{figure*}

As mentioned in the previous section, we introduced the cutoff parameters $\Lambda_\pex$ and $\Lambda_\vex$ as additional parameters.
While these parameters are independent of the DCC model employed to describe the $\mex \bi \to \mx\bx$ subprocesses and 
should ultimately be determined through a comprehensive analysis of both the $\mi \bi \to \mf \mx\bx$ and the $\bar K N$ reactions, 
we tentatively fixed them at $1$~GeV in the present study.
To examine the validity of these chosen values, 
we show in Fig.~\ref{fig:pimp2KsSs} a comparison of $d\sigma/dt$ with 
the old data for the $\pi^- p \to K^{*0} \Sigma^{*0}(1385)$ reaction at $|\moms{p}_\mi^\lfr|=3.95$~GeV/$c$~\cite{CERN-CFMS1980}.
(Here, the three-momenta labeled with the superscript ``$\lfr$'' denote those in the laboratory frame.)
Our results for the $\pi^- p \to K^{*0}\Sigma^{*0}(1385)$ reaction are obtained as follows:
Using Eqs.~(\ref{eq:crs-dtdwdo})-(\ref{eq:mimfvex-vtx}),
we compute $d\sigma/(dtdW)$ for the $\pi^- p \to K^{*0}\pi^0 \Lambda$ reaction
and perform an integration over $W$ (the $\pi^0\Lambda$ invariant mass) in the range $W = 1.34$-$1.44$~GeV,
consistent with the analysis in Ref.~\cite{CERN-CFMS1980}.
We then assume that the $\pi^0\Lambda$ originates from the decay of
the $\Sigma^{*0}(1385)$ via the two-step process $\pi^- p \to K^{*0}\Sigma^{*0}(1385)\to K^{*0}\pi^0 \Lambda$.
Because the $\Sigma^{*0}(1385)$ can also decay through other channels, we
divide the integrated cross section 
by the branching fraction of the $\Sigma^{*0}(1385) \to \pi^0 \Lambda$ decay,
${\cal B}[\Sigma^{*0}(1385) \to \pi^0 \Lambda] \approx 0.87$~\cite{PDG2024}.
In Fig.~\ref{fig:pimp2KsSs},
we present the results calculated using Model~A and Model~B of the DCC model in Ref.~\cite{Kamano2014} 
to evaluate the $\mex N \to \mx \bx$ subprocesses.
[Hereafter, we simply refer to the calculations for the $\mi \bi \to \mf \mx \bx$ reaction using 
Model~A (Model~B) of Ref.~\cite{Kamano2014} as Model~A (Model~B).]
The results are plotted as a function of $-t'$, where $t'$ is defined as $t'\equiv t-t_{\textrm{max}}$,
with $t_{\textrm{max}}$ being the maximum value of $t$ for a given $W$, $|\moms{p}_\mi^\lfr|$, and $M_\mf$.
At $t=t_\textrm{max}$, the scattering angle of the outgoing $\mf$ in the laboratory frame, $\theta_\mf^\lfr$,
becomes zero.
We see that both Model~A and Model~B reproduce the data reasonably well up to $-t' \approx 1$~GeV$^2$,
indicating that the chosen values for $\Lambda_\pex$ and $\Lambda_\vex$ are valid.
In both models, the $\pex$-exchange process dominates the cross section in the very forward $\mf$ production region at $t'\approx 0$.
However, the behavior of the $\vex$-exchange process differs significantly between the two models.
In Model~A, the $\vex$-exchange process has a non-negligible contribution even at $t'\approx 0$ and 
becomes larger than the $\pex$-exchange process at $-t' > 0.2$~GeV$^2$.
On the other hand, the $\vex$-exchange process in Model~B shows a negligible contribution for this reaction at the considered energy.

With the parameters $\Lambda_\pex$ and $\Lambda_\vex$ of our model fixed as explained above, 
we can proceed to make predictions of the cross sections for the $\pi^{\pm}p \to \mf\mx\bx$ reactions.
To identify the energy region where $Y^*$ resonances can be studied, 
we first present the results for $d\sigma/(dtdW)$.
We then examine the partial-wave contributions to $d\sigma/(dtdW)$ 
to discuss the effects of the $Y^*$ resonances on the angular distributions $d\sigma/(dtdWd\Omega^*_\mx)$.

\subsection{$\pi^- p \to \mf^0 \mx \bx$}

We now investigate the $\pi^- p \to \mf^0 \mx \bx$ reaction 
at an incident beam momentum of $|\moms{p}_\mi^\lfr| = 10$~GeV/$c$,
where the highest intensity of the $\pi^-$ beam is expected at J-PARC~\cite{Noumi}.

\subsubsection{$d\sigma/(dtdW)$ up to $W=2.1$~GeV and pronounced $Y^*$s}

Figure~\ref{fig:dsdtdw-all} shows $d\sigma/(dtdW)$ 
at $|\moms{p}_\mi^\lfr| = 10$~GeV/$c$ and $t'=0$
for the $\pi^- p \to K^{*0} \mx \bx$ reactions 
with various final $\mx\bx$ states: 
$\mx\bx = \pi^0\Lambda$, $\pi^-\Sigma^+$, $\pi^0\Sigma^0$, $\pi^+\Sigma^-$, $K^- p$, $\bar K^0 n$, $\eta \Lambda$, $K^0\Xi^0$, and $K^+\Xi^-$.
The results are presented as a function of $W$ from the threshold of each channel up to $2.1$~GeV.
In the $W$ region above the $\bar K N$ threshold, a clear peak observed at $W\approx 1.52$~GeV 
in the $\mx\bx =\pi\Sigma$ and $\bar K N$ channels is attributed to the well-known $\Lambda(1520)3/2^-$ resonance.
For the $\eta\Lambda$ channel, the sharp rise of the cross section and the peak structure observed at the threshold 
are attributed to the $S$-wave $\Lambda(1670)1/2^-$ resonance.
(Note that in Model~B, an unconfirmed narrow $P_{03}$ $\Lambda$ resonance with a pole mass of $1671-i5$~MeV 
also contributes to this structure, as discussed in Ref.~\cite{Kamano2015}.)
In the $W\approx 1.8$~GeV region, although a number of $Y^*$ resonances have been reported, 
only a single bump is observed in the cross sections for most of the $\mx\bx$ channels presented.
This reflects the nature of light-baryon resonances, which are in general broad and highly overlapping.
This is the reason why a detailed partial-wave analysis of angular distributions and polarization observables 
is necessary for establishing the light-baryon spectrum.
In the same figures, we also present the individual contributions of the $\pex$-exchange and $\vex$-exchange processes.
For the $\mx\bx = \pi\Lambda$, $\pi\Sigma$, and $\bar K N$ channels,
the contribution of the $\pex$-exchange process generally dominates the cross sections in the presented $W$ region.
The $\vex$-exchange contribution becomes significant or comparable to that of the $\pex$-exchange process at high $W$ values
for the $\pi\Lambda$ and $\pi\Sigma$ channels, while it remains negligible for the $\bar K N$ channel.
On the other hand, the situation is reversed for the $\mx\bx=\eta\Lambda$ and $K\Xi$ channels,
where the contribution of the $\vex$-exchange process is dominant or comparable to 
that of the $\pex$-exchange process even near the thresholds.

For the $\mx\bx = \pi \Lambda$ and $\pi\Sigma$ channels in Fig~\ref{fig:dsdtdw-all},
large cross sections are observed in the $W$ region below the $\bar K N$ threshold,
indicating that these reactions will indeed provide critical information to investigate $Y^*$ resonances
located in this region, which are not directly accessible via $\bar K N$ reactions.
As presented in Sec.~\ref{sec:low-ystar}, 
within our 2014 DCC analysis of $\bar K N$ reactions~\cite{Kamano2014},
two $3/2^+$ $\Sigma$ resonances were found
below the $\bar K N$ threshold in the $P_{13}$ partial wave for both Model~A and Model~B.
One of them is the well-established $\Sigma^*(1385)3/2^+$ resonance with a pole mass of $1381-i20$~MeV,
while the other may be a new resonance.
For the $\mx\bx=\pi^0\Lambda$ channel, a clear peak and/or bump structure 
is observed below the $\bar K N$ threshold.
In Model~A, a distinct peak and a bump are observed at $W\approx 1.31$~GeV and $W\approx 1.38$~GeV, respectively.
Since in this $W$ region the contribution from the $P_{13}$ partial wave in
the $\mex N \to \pi^0 \Lambda$ subprocess dominates the cross section
(see Fig.~\ref{fig:dsdtdw-pwa-piL}), the peak (bump) can be attributed to
the new $3/2^+$ $\Sigma$ resonance with a pole mass of $1303-i13$~MeV 
[the well-established $\Sigma^*(1385)3/2^+$ with a pole mass of $1381-i20$~MeV].
On the other hand, only a single peak is observed at $W\approx 1.37$~GeV in Model~B,
although two $\Sigma^*(3/2^+)$ resonances are found to exist also in this model.
This is likely because the pole positions of these two resonances,
$1360-i38$~MeV for the new $3/2^+$ $\Sigma$ resonance and $1381-i20$~MeV for the well-established $\Sigma^*(1385)3/2^+$,
are closer to each other compared to the case of Model~A;
furthermore, the pole of the new $3/2^+$ $\Sigma$ resonance is located further from the real energy axis 
than the $\Sigma^*(1385)3/2^+$ pole, making it less visible in the cross section.
Indeed, a distinct peak in the $\pi^0\Lambda$ spectrum at $W\approx 1.31$~GeV, 
as presented in Model~A, has never been reported experimentally.
However, given the results of Model~B, this fact does not exclude the possibility of 
the existence of two $3/2^+$ $\Sigma$ resonances below the $\bar K N$ threshold,
since two resonances can manifest as a single peak in the cross section.
In fact, this situation is analogous to the case of the $1/2^-$ $\Lambda$ resonances below the $\bar K N$ threshold.
Historically, the experimental observation of a single peak in the $\pi\Sigma$ spectrum led to the assumption of a single resonance.
(We note that such a single-peak structure is also observed in our results for the $\mx\bx = \pi^-\Sigma^+, \pi^0\Sigma^0,$ and $\pi^+\Sigma^-$ channels.)
However, it is now widely recognized that two $1/2^-$ $\Lambda$ resonances coexist in this energy region~\cite{PDG2024}.
High-statistics data for the $\pi N \to \mf \mx \bx$ reactions are highly desirable
to clarify the full picture of the $3/2^+$ $\Sigma$ resonances
in the $P_{13}$ partial wave below the $\bar K N$ threshold.

\subsubsection{Partial-wave contributions}

\begin{figure}[tb]
\centering
\includegraphics[clip,width=0.48\textwidth]{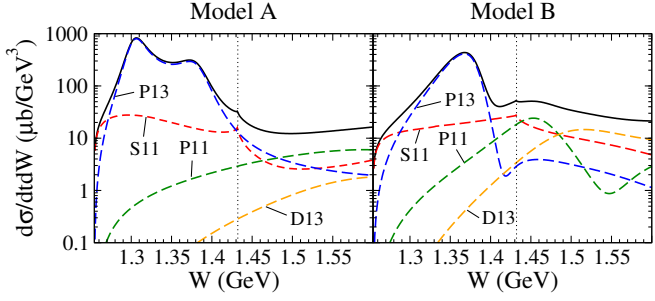}
\caption{
Contributions of individual partial waves in the $\mex \bi \to \mx\bx$ subprocess 
to $d\sigma/(dtdW)$ for the $\pi^- p \to \mf^0 \pi^0 \Lambda$ reaction at $|\moms{p}_\mi^\lfr| = 10$~GeV/$c$ and $t'=0$.
Only partial-wave contributions with $J \leq 3/2$ are shown for the $W$ range from the threshold up to $W=1.6$~GeV.
The left (right) panel shows the results for Model~A (Model~B).
The solid curves represent the full results, while the dashed curves correspond to those obtained 
by retaining only the specific single partial wave indicated in the figure when constructing the invariant amplitude for the $\mex \bi \to \mx\bx$ subprocess.
The vertical dotted line indicates the $\bar K N$ threshold.
}
\label{fig:dsdtdw-pwa-piL}
\end{figure}
\begin{figure}[tb]
\centering
\includegraphics[clip,width=0.48\textwidth]{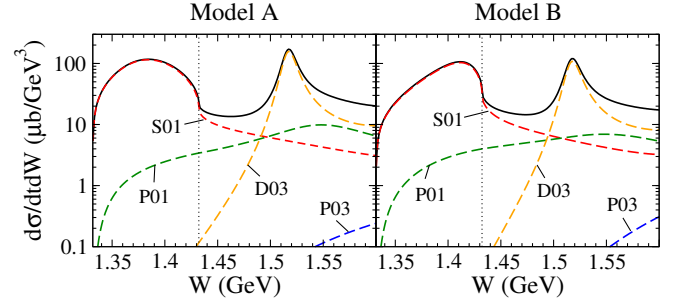}
\caption{
Contributions of individual partial waves in the $\mex \bi \to \mx\bx$ subprocess 
to $d\sigma/(dtdW)$ for the $\pi^- p \to \mf^0 \pi^0 \Sigma^0$ reaction at $|\moms{p}_\mi^\lfr| = 10$~GeV/$c$ and $t'=0$.
Only partial-wave contributions with $J \leq 3/2$ are shown for the $W$ range from the threshold up to $W=1.6$~GeV.
The left (right) panel shows the results for Model~A (Model~B).
The solid curves represent the full results, while the dashed curves correspond to those obtained 
by retaining only the specific single partial wave indicated in the figure when constructing the invariant amplitude for the $\mex \bi \to \mx\bx$ subprocess.
The vertical dotted line indicates the $\bar K N$ threshold.
}
\label{fig:dsdtdw-pwa-pi0S0}
\end{figure}
\begin{figure}[tb]
\centering
\includegraphics[clip,width=0.48\textwidth]{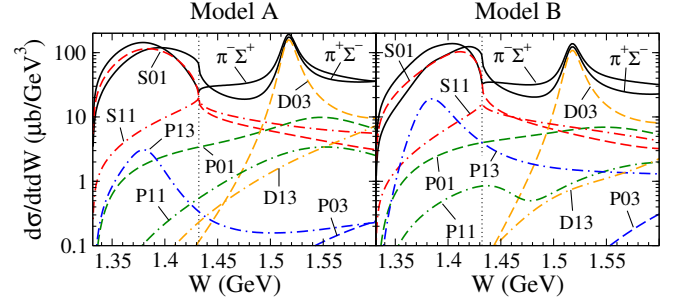}
\caption{
Contributions of individual partial waves in the $\mex \bi \to \mx\bx$ subprocess 
to $d\sigma/(dtdW)$ for the $\pi^- p \to \mf^0 \pi^- \Sigma^+$ and $\pi^- p \to \mf^0 \pi^+ \Sigma^-$ reactions at $|\moms{p}_\mi^\lfr| = 10$~GeV/$c$ and $t'=0$.
Only partial-wave contributions with $J \leq 3/2$ are shown for the $W$ range from the threshold up to $W=1.6$~GeV.
The left (right) panel shows the results for Model~A (Model~B).
The solid curves represent the full results for the $\mx\bx=\pi^-\Sigma^+$ and $\pi^+\Sigma^-$ channels, while the dashed and dot-dashed curves correspond to those obtained 
by retaining only the specific single partial wave when constructing the invariant amplitude for the $\mex \bi \to \mx\bx$ subprocess.
The results that each curve represents are indicated in the figure. 
The vertical dotted line indicates the $\bar K N$ threshold.
}
\label{fig:dsdtdw-pwa-piS-0c}
\end{figure}

To understand the structure of $d\sigma/(dtdW)$ for the $\mx\bx = \pi^0\Lambda$ channel shown in Fig.~\ref{fig:dsdtdw-all},
we present in Fig.~\ref{fig:dsdtdw-pwa-piL} the contributions of individual partial waves in the $\mex \bi \to \mx\bx$ subprocess 
to $d\sigma/(dtdW)$ for the $\pi^- p \to \mf^0 \pi^0 \Lambda$ reaction. 
These contributions are obtained by retaining only the corresponding partial wave when constructing the invariant amplitude for the subprocess.

For the $\mx\bx=\pi^0\Lambda$ channel, only the isospin $I=1$ partial waves contribute.
The $S_{11}$ contribution is dominant in the $W$ region close to the threshold; subsequently, the $P_{13}$ contribution dominates the cross section up to $W\approx 1.4$~GeV. 
Above $W\approx 1.4$~GeV, the $S_{11}$, $P_{11}$, and $P_{13}$ contributions become comparable in Model~A, whereas the $S_{11}$, $P_{11}$, and $D_{13}$ contributions become comparable in Model~B.
In the energy region $W \leq 1.6$~GeV, contributions from higher partial waves with $J \geq 5/2$ 
are found to be very minor or negligible and are thus omitted here. 
This trend also holds for other $\pi p \to \mf \mx \bx$ reactions discussed in this paper; accordingly, 
only partial-wave contributions up to $J = 3/2$ will be presented in subsequent similar figures for consistency.
We also see in Fig.~\ref{fig:dsdtdw-pwa-piL} that the $W$ dependencies of the $P_{11}$ and $D_{13}$ contributions differ significantly between the two models.
While Model~A shows monotonically increasing behavior for both the $P_{11}$ and $D_{13}$ contributions as $W$ increases, 
Model~B exhibits a peak at $W\approx 1.45$~GeV for the $P_{11}$ contribution and a bump at $W\approx 1.5$~GeV for the $D_{13}$ contribution.
This distinct difference is likely attributable to the existence of low-lying $P_{11}$ and $D_{13}$ resonances present in Model~B.
As presented in Table~\ref{tab:low-ystar}, Model~A exhibits the $1/2^-$ $\Sigma$ resonance with a pole mass of $1381-i150$~MeV, 
which is in the same energy region as the $\Sigma(1385)3/2^+$.
Due to the large imaginary part of its pole mass, this resonance does not present a clear peak or bump in the $S_{11}$ contribution.
Although the $S_{11}$ contribution is the second largest in the $\Sigma(1385)3/2^+$ region, 
it is overshadowed by the dominant $P_{13}$ contribution; 
thus, it will be difficult to identify this $1/2^-$ $\Sigma$ resonance 
solely by examining the $W$ dependence of the cross section.

In Fig.~\ref{fig:dsdtdw-pwa-pi0S0}, we present the contributions of individual partial waves in the $\mex \bi \to \mx\bx$ subprocess 
to $d\sigma/(dtdW)$ for the $\pi^- p \to \mf^0 \pi^0 \Sigma^0$ reaction.
For the $\mx\bx=\pi^0\Sigma^0$ channel, only the isospin $I=0$ partial waves contribute.
We find that the cross section is completely dominated by the $S_{01}$ contribution below the $\bar K N$ threshold, 
while the $P_{01}$ contribution also becomes significant above the threshold.
The peak at $W\approx 1.52$~GeV arising from the $D_{03}$ contribution is a direct manifestation of the $\Lambda(1520)3/2^-$ resonance.
The overall behaviors of the individual partial-wave contributions do not differ significantly between Model~A and Model~B, 
except for the line shape of the $S_{01}$ contribution below the $\bar K N$ threshold. 
This is attributable to the difference in the pole positions of the two $1/2^-$ $\Lambda$ resonances,
and is clearly manifested in the $W$ distribution.
A similar behavior has also been observed in the $\pi^0\Sigma^0$ mass distributions for the $K^- d \to \pi^0\Sigma^0 n$ reaction,
as discussed in Ref.~\cite{Kamano2016}.
Although Model~B exhibits a $1/2^+$ $\Lambda$ resonance with a pole mass of $1401-i155$~MeV (see Table~\ref{tab:low-ystar}), 
this resonance does not present a clear peak or bump in the $P_{01}$ contribution. 
This is likely attributable to the large imaginary part of its pole mass, similar to the case of the $1/2^-$ $\Sigma$ resonance found in Model~A.

Fig.~\ref{fig:dsdtdw-pwa-piS-0c} shows the contributions of individual partial waves in the $\mex \bi \to \mx\bx$ subprocess
to $d\sigma/(dtdW)$ for the $\pi^- p \to \mf^0 \pi^- \Sigma^+$ and $\pi^- p \to \mf^0 \pi^+ \Sigma^-$ reactions.
Assuming isospin symmetry, the contribution of each individual partial wave is identical for both the $\pi^-\Sigma^+$ and $\pi^+\Sigma^-$ channels, 
as they differ only in their isospin Clebsch-Gordan coefficients.
Therefore, we present them together in Fig.~\ref{fig:dsdtdw-pwa-piS-0c}, 
where the dashed and dot-dashed curves indicate the $I=0$ and $I=1$ partial waves, respectively.
Below the $\bar K N$ threshold, the $S_{01}$ contribution is dominant, similar to the $\mx\bx=\pi^0\Sigma^0$ case (Fig.~\ref{fig:dsdtdw-pwa-pi0S0}), 
followed by the $S_{11}$ and $P_{13}$ contributions.
Notably, the $P_{13}$ contribution is overshadowed by the dominant $S_{01}$ term; consequently, 
a clear peak structure in the $\Sigma(1385)3/2^+$ region, such as that seen in the $\mx\bx=\pi^0\Lambda$ channel (Fig.~\ref{fig:dsdtdw-pwa-piL}), is not observed here.
As with Model~B, only a single peak is observed in the $P_{13}$ contribution in Model~A for these reactions.
This is expected because the new $3/2^+$ $\Sigma$ resonance found in Model~A, with a pole mass of $1303-i13$~MeV, 
is located below the $\pi\Sigma$ threshold; thus, its contribution to the cross section is kinematically suppressed.
The general trend that the $I=1$ contributions are subdominant and obscured by the $I=0$ partial waves persists above the $\bar{K}N$ threshold. 
This dominance of the $I=0$ components could make it difficult to investigate the low-lying $1/2^+$ and $3/2^-$ $\Sigma$ resonances found in Model B using these reactions.
In fact, the contributions from $P_{11}$ and $D_{13}$ partial waves are rather small compared with dominant partial waves.

\subsubsection{Angular distributions}

To perform a detailed partial-wave analysis and reliably extract the $Y^*$ resonance parameters 
from the $\mi \bi \to \mf \mx \bx$ reaction data, 
relying solely on $d\sigma/dt$ and $d\sigma/(dtdW)$ is insufficient.
It is necessary to examine $Y^*$ contributions to the angular distributions.
We therefore turn our attention to investigating the angular distributions of the final $\mx\bx$ states 
at given $t$ and $W$, namely $d\sigma/(dtdWd\Omega^*_\mx)$.

\begin{figure}[tb]
\centering
\includegraphics[clip,width=0.48\textwidth]{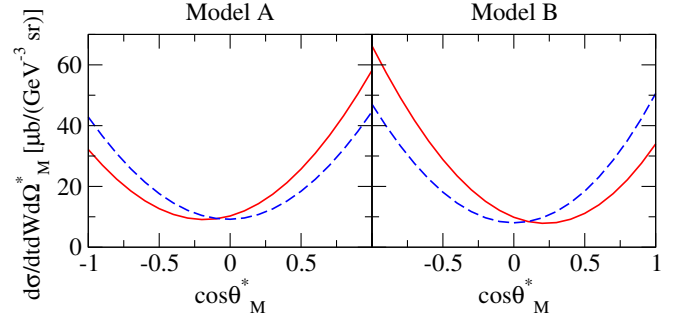}
\caption{
The $\cos\theta^*_\mx$ dependence of $d\sigma/(dtdWd\Omega^*_\mx)$ 
for the $\pi^- p \to \mf^0 \pi^0 \Lambda$ reaction at $|\moms{p}_\mi^\lfr| = 10$~GeV/$c$, $t'=0$, and $W=1381$~MeV. 
The azimuthal angle of the meson $\mx$ ($\phi^*_\mx$) is set to zero.
The left (right) panel shows the results for Model~A (Model~B).
The red solid curves represent the full results, while the blue dashed curves 
show the results obtained by excluding the contribution of the $S_{11}$ partial wave 
in the $\mex \bi \to \mx \bx$ subprocess.
}
\label{fig:dsdtdwdO-piL-1381mev}
\end{figure}
\begin{figure}[tb]
\centering
\includegraphics[clip,width=0.48\textwidth]{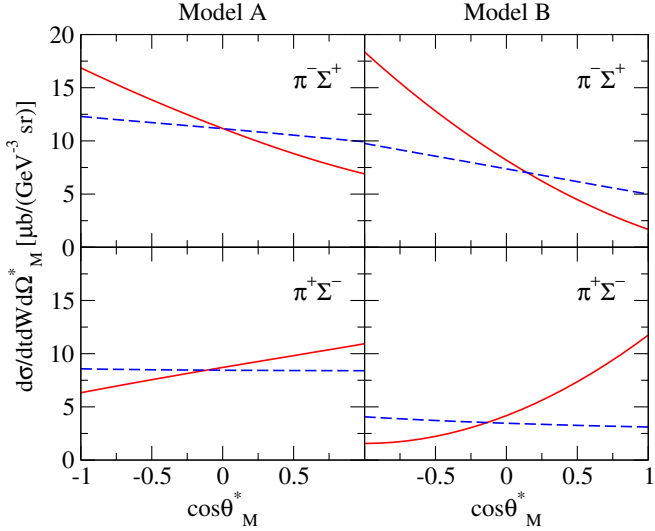}
\caption{
The $\cos\theta^*_\mx$ dependence of $d\sigma/(dtdWd\Omega^*_\mx)$ 
for the $\pi^- p \to \mf^0 \pi^- \Sigma^+$ and $\pi^- p \to \mf^0 \pi^+ \Sigma^-$
reactions at $|\moms{p}_\mi^\lfr| = 10$~GeV/$c$, $t'=0$, and $W=1381$~MeV. 
The azimuthal angle of the meson $\mx$ ($\phi^*_\mx$) is set to zero.
The red solid curves represent the full results, while the blue dashed curves 
show the results obtained by excluding the contribution of the $P_{13}$ partial wave 
in the $\mex \bi \to \mx \bx$ subprocess.
}
\label{fig:dsdtdwdO-piS-1381mev}
\end{figure}
\begin{figure}[tb]
\centering
\includegraphics[clip,width=0.48\textwidth]{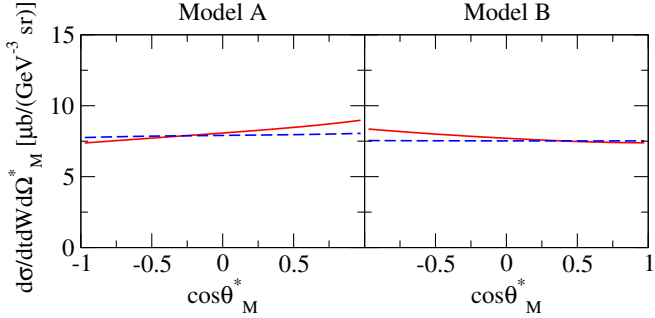}
\caption{
The $\cos\theta^*_\mx$ dependence of $d\sigma/(dtdWd\Omega^*_\mx)$ 
for the $\pi^- p \to \mf^0 \pi^0 \Sigma^0$ reaction 
at $|\moms{p}_\mi^\lfr| = 10$~GeV/$c$, $t'=0$, and $W=1401$~MeV. 
The azimuthal angle of the meson $\mx$ ($\phi^*_\mx$) is set to zero.
The red solid curves represent the full results, while the blue dashed curves 
show the results obtained by excluding the contribution of the $P_{01}$ partial wave
in the $\mex \bi \to \mx \bx$ subprocess.
}
\label{fig:dsdtdwdO-piS-1401mev}
\end{figure}
\begin{figure}[tb]
\centering
\includegraphics[clip,width=0.48\textwidth]{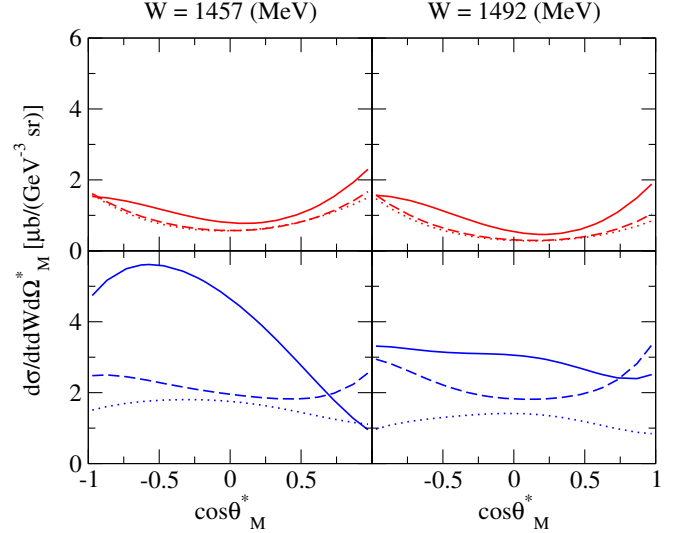}
\caption{
The $\cos\theta^*_\mx$ dependence of $d\sigma/(dtdWd\Omega^*_\mx)$ 
for the $\pi^- p \to \mf^0 \pi^0 \Lambda$ reaction 
at $|\moms{p}_\mi^\lfr| = 10$~GeV/$c$, $t'=0$. 
The results at $W=1457$ and $1492$~MeV are presented.
The azimuthal angle of the meson $\mx$ ($\phi^*_\mx$) is set to zero.
The top (bottom) panels show the results of Model~A (Model~B).
The solid curves represent the full results, while the dashed (dotted) curves 
show the results obtained by excluding the contribution of the $P_{11}$ partial wave
(the contributions of the $P_{11}$ and $D_{13}$ partial waves)
in the $\mex \bi \to \mx \bx$ subprocess.
}
\label{fig:dsdtdwdO-piL-1457-1492mev}
\end{figure}

Figure~\ref{fig:dsdtdwdO-piL-1381mev} shows the $\cos\theta^*_\mx$ dependence of $d\sigma/(dtdWd\Omega^*_\mx)$ 
for the $\pi^- p \to \mf^0 \pi^0 \Lambda$ reaction at $|\moms{p}_\mi^\lfr| = 10$~GeV/$c$, $t'=0$, and $W=1381$~MeV. 
Since there is no $\phi_\mx^*$ dependence for $d\sigma/(dtdWd\Omega^*_\mx)$ at $t'=0$, we simply set $\phi_\mx^*=0$.
Here, $W=1381$~MeV corresponds to the real part of the pole mass of the well-established $\Sigma(1385)3/2^+$; 
thus, the $P_{13}$ contribution completely dominates $d\sigma/(dtdW)$ as shown in Fig.~\ref{fig:dsdtdw-pwa-piL}, 
followed by the $S_{11}$ contribution as the subdominant one.
We also note that the $\vex$-exchange process has a negligible contribution in this $W$ region,
as can be seen from Fig.~\ref{fig:dsdtdw-all}.
In this situation, the $\cos\theta_\mx^*$ dependence of the unpolarized cross section can qualitatively take the form:
$|{\cal A}_{S_{11}}|^2 + ({\cal A}_{S_{11}}{\cal A}_{P_{13}}^* + \textrm{c.c.})\cos\theta_\mx^* + |{\cal A}_{P_{13}}|^2 (3\cos^2\theta_\mx^* +1) +\cdots$,
where ${\cal A}_{S_{11}}$ (${\cal A}_{P_{13}}$) generically denotes the quantity encompassing the amplitude and other relevant factors associated with the $S_{11}$ ($P_{13}$) contribution 
of the $\pex$-exchange process, with its specific $\cos\theta_\mx^*$ dependence factored out in each respective term, and 
the terms involving either the $\vex$-exchange process or any partial wave other than $S_{11}$ and $P_{13}$ are suppressed.
The angular distribution shown in Fig.~\ref{fig:dsdtdwdO-piL-1381mev} exhibits an upward-opening parabolic shape because the term proportional to $\cos^2\theta_\mx^*$ 
becomes dominant due to the large $P_{13}$ contribution.
Furthermore, the asymmetric behavior of the full results with respect to $\cos\theta_\mx^* = 0$ originates from 
the term proportional to $\cos\theta_\mx^*$, which arises from the interference between the dominant $P_{13}$ and subdominant $S_{11}$ contributions.
In fact, when the $S_{11}$ contribution is turned off, the angular distribution becomes symmetric around $\cos\theta_\mx^* = 0$.
This indicates that the subdominant $S_{11}$ contribution is amplified by the interference with the dominant $P_{13}$ contribution, 
thereby exhibiting a clearly visible effect on the angular distribution.
Consequently, the angular distribution in this $W$ region can provide crucial information to validate or refute the existence of another low-lying $3/2^+$ $\Sigma$ resonance 
found in both Model~A and Model~B, as well as the $1/2^-$ $\Sigma$ resonance obtained in Model~A, 
both of which are potentially new resonances.

Figure~\ref{fig:dsdtdwdO-piS-1381mev} shows the $\cos\theta^*_\mx$ dependence of $d\sigma/(dtdWd\Omega^*_\mx)$ 
for the $\pi^- p \to \mf^0 \pi^- \Sigma^+$ and $\pi^- p \to \mf^0 \pi^+ \Sigma^-$ reactions 
at $|\moms{p}_\mi^\lfr| = 10$~GeV/$c$, $t'=0$, and $W=1381$~MeV. 
As can be seen from Fig.~\ref{fig:dsdtdw-pwa-piS-0c}, in the energy region around $W=1381$~MeV, 
the $S_{01}$ contribution is dominant, while the effect of the subdominant $P_{13}$ contribution is negligible and difficult to observe in the $W$ dependence of $d\sigma/(dtdW)$.
However, the effect of this subdominant $P_{13}$ contribution is clearly visible in the angular distribution.
In fact, the linearly increasing (decreasing) behavior of the full results with respect to $\cos\theta^*_\mx$ 
for the $\mx\bx=\pi^-\Sigma^+$ ($\mx\bx=\pi^+\Sigma^-$) case mostly originates from the effect of the subdominant $P_{13}$ contribution.
Similar to the discussion regarding Fig.~\ref{fig:dsdtdwdO-piL-1381mev}, 
this can be understood from the $\cos\theta_\mx^*$ dependence of the cross section arising from 
the major partial-wave contributions in the $\mex \bi \to \mx \bx$ subprocess, which qualitatively takes the form:
$|{\cal A}_{S_{01}}|^2 + ({\cal A}_{S_{01}}{\cal A}_{P_{13}}^* + \textrm{c.c.})\cos\theta_\mx^* + |{\cal A}_{P_{13}}|^2 (3\cos^2\theta_\mx^* +1) +\cdots$,
where ${\cal A}_{S_{01}}$ (${\cal A}_{P_{13}}$) generically denotes the quantity encompassing the amplitude and other relevant factors associated with the $S_{01}$ ($P_{13}$) contribution 
of the $\pex$-exchange process, with its specific $\cos\theta_\mx^*$ dependence factored out in each respective term, and 
the minor terms involving either the $\vex$-exchange process or any partial wave other than $S_{01}$ and $P_{13}$ are suppressed.
In this case, the linear $\cos\theta^*_\mx$ dependence of the angular distribution results from the subdominant $P_{13}$ contribution amplified by the interference with the dominant $S_{01}$ contribution.
Indeed, the results excluding the subdominant $P_{13}$ contribution (blue dashed curves in Fig.~\ref{fig:dsdtdwdO-piS-1381mev}) show a rather flat angular dependence, 
which is characteristic of the $S$-wave contribution.
Additionally, the slightly concave-up behavior of the full results in Model~B is attributable to the $|{\cal A}_{P_{13}}|^2 \cos^2\theta_\mx^*$ term, 
as the $P_{13}$ contribution in Model~B is larger than that in Model~A.
It should also be noted that the different linear dependence between 
the $\mx\bx=\pi^-\Sigma^+$ and $\mx\bx=\pi^+\Sigma^-$ cases, where the former (latter) shows 
linearly increasing (decreasing) behavior,
is attributable to the sign difference in the isospin Clebsch-Gordan coefficients for the $I=1$ $\pi\Sigma$ state.
Specifically, the isospin $I=1$ partial waves are multiplied by 
$\cg{I_\pi I_\pi^z}{I_\Sigma I_\Sigma^z}{II^z} = \cg{1-1}{1+1}{10}=-1/\sqrt{2}$ for the $\mx\bx=\pi^-\Sigma^+$ case, 
whereas they are multiplied by 
$\cg{I_\pi I_\pi^z}{I_\Sigma I_\Sigma^z}{II^z} = \cg{1+1}{1-1}{10}=+1/\sqrt{2}$ for the $\mx\bx=\pi^+\Sigma^-$ case.

Figure~\ref{fig:dsdtdwdO-piS-1401mev} shows the $\cos\theta^*_\mx$ dependence of $d\sigma/(dtdWd\Omega^*_\mx)$ 
for the $\pi^- p \to \mf^0 \pi^0 \Sigma^0$ reaction at $|\moms{p}_\mi^\lfr| = 10$~GeV/$c$, $t'=0$, and $W=1401$~MeV. 
This $W$ value corresponds to the real part of the pole mass of the low-lying $1/2^+$ $\Lambda$ resonance found in Model~B.
Within the predictions of our current models, unlike the cases shown in Figs.~\ref{fig:dsdtdwdO-piL-1381mev} and~\ref{fig:dsdtdwdO-piS-1381mev}, 
the enhancement of the subdominant $P_{01}$ contribution through interference with the dominant $S_{01}$ contribution is found to be very small, 
although it produces a slight linear dependence on $\cos\theta^*_\mx$.
To validate or refute the existence of the low-lying $1/2^+$ $\Lambda$ resonance found in Model~B, 
measurements of the angular distribution in this energy region are highly desirable.
If future experimental data exhibit a behavior strongly dependent on $\cos\theta^*_\mx$ or $\cos^2\theta^*_\mx$, 
the situation could change significantly.

Figure~\ref{fig:dsdtdwdO-piL-1457-1492mev} shows the $\cos\theta^*_\mx$ dependence of $d\sigma/(dtdWd\Omega^*_\mx)$ 
for the $\pi^- p \to \mf^0 \pi^0 \Lambda$ reaction at $|\moms{p}_\mi^\lfr| = 10$~GeV/$c$ and $t'=0$.
The results at $W=1457$ and $1492$~MeV are presented.
The former (latter) $W$ value corresponds to the real part of the pole mass of the low-lying $1/2^+$ ($3/2^-$) $\Sigma$ resonance found in Model~B.
In Model~B, the behavior of the angular distribution changes drastically depending on the $W$ value in this $W$ region.
This variation is primarily driven by the $P_{11}$ and $D_{13}$ partial waves, 
which contain the low-lying resonance states in this $W$ region.
Indeed, when the contributions from both of these partial waves are turned off (dotted curves), 
the angular behaviors at the two $W$ values become almost identical.
In contrast, the results from Model~A show little variation in both the magnitude and shape of the angular distribution 
across this $W$ region, where the contributions from the $P_{11}$ and $D_{13}$ partial waves are relatively small.
This distinct difference in the angular dependence between Model~A and Model~B 
suggests that measuring the angular distribution in this $W$ region would be highly effective 
for validating or refuting the existence of the low-lying $\Sigma$ resonances found in Model~B.
Therefore, experimental measurements in this specific kinematic region are strongly desired.

\subsection{$\pi^+ p \to \mf^+ \mx \bx$}

We now turn to the investigation of the $\pi^+ p \to \mf^+ \mx \bx$ reaction.
The primary advantage of this reaction is that the two-body $\mx\bx$ state 
has a total charge of $+1$, and therefore cannot contain any total isospin $I=0$ contributions, which include the $\Lambda$ resonances.
Consequently, this reaction allows us to exclusively investigate the isospin $I=1$ $\Sigma$ resonances.

As in the previous section, we first present the results for $d\sigma/(dtdW)$ and then examine the partial-wave contributions 
to $d\sigma/(dtdW)$ to discuss the effects of the $Y^*$ resonances on the angular distributions $d\sigma/(dtdWd\Omega^*_\mx)$.

\subsubsection{$d\sigma/(dtdW)$ up to $W=2.1$~GeV and pronounced $Y^*$s}

\begin{figure*}[tb]
\centering
\includegraphics[clip,width=0.8\textwidth]{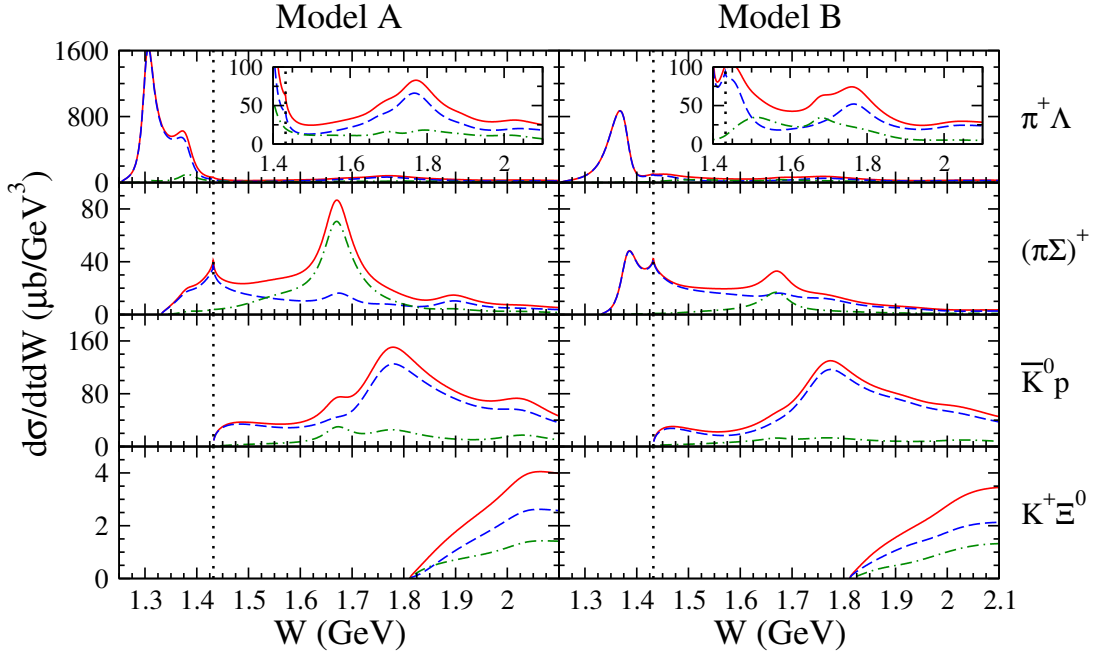}
\caption{
Differential cross section $d\sigma/(dtdW)$ for the $\pi^+ p \to K^{*+} \mx \bx$ reactions at $|\moms{p}_\mi^\lfr| = 10$~GeV/$c$ and $t'=0$.
The left (right) panels show the results from Model~A (Model~B).
Each row corresponds to a specific $\mx\bx$ state, as indicated on the right side of the row.
Note that under isospin symmetry, the cross sections for the $\mx\bx = \pi^+\Sigma^0$ and $\pi^0\Sigma^+$ channels are identical.
Therefore, these channels are presented collectively under the notation $\mx\bx=(\pi\Sigma)^+$, which represents the results for either state.
The insets in the $\mx\bx=\pi^+\Lambda$ panels show an enlarged view of the $W\geq 1.4$~GeV region.
The meaning of each curve is the same as in Fig.~\ref{fig:pimp2KsSs}.
The dotted vertical line in each panel represents the $\bar{K}N$ threshold energy.
}
\label{fig:dsdtdw-all-plus}
\end{figure*}

Figure~\ref{fig:dsdtdw-all-plus} shows $d\sigma/(dtdW)$ at $|\moms{p}_\mi^\lfr| = 10$~GeV/$c$ and $t'=0$
for the $\pi^+ p \to K^{*+} \mx \bx$ reactions with various final $\mx\bx$ states: 
$\mx\bx = \pi^+\Lambda$, $\pi^+\Sigma^0$, $\pi^0\Sigma^+$, $\bar{K}^0 p$, and $K^+\Xi^0$.
The results are presented as a function of $W$ from the threshold of each channel up to 2.1~GeV.
Under isospin symmetry, the cross sections for the $\mx\bx = \pi^+\Sigma^0$ and $\pi^0\Sigma^+$ channels are identical. 
Therefore, rather than displaying the results for both channels separately, 
we present them collectively in the figure under the notation $\mx\bx=(\pi\Sigma)^+$, 
which represents the results for either of these $\mx\bx$ states.
For the $\mx\bx = \pi\Lambda$ channels, 
the magnitude of $d\sigma/(dtdW)$ for the $\pi^+ p \to K^{*+} \pi^+\Lambda$ reaction is found to be exactly twice 
that for the $\pi^- p \to K^{*0} \pi^0\Lambda$ reaction (compare with the results for $\mx\bx=\pi^0\Lambda$ in Fig.~\ref{fig:dsdtdw-all}).
We note that this exact factor of two is a direct consequence of isospin symmetry combined with 
the $t$-channel one-meson exchange mechanism considered in this work.
Therefore, experimentally measuring and comparing the cross sections of these two reactions 
would provide an excellent test of whether the assumption of $t$-channel one-meson exchange dominance 
is valid at the considered beam momentum and kinematic region.

For the $\mx\bx= \pi\Sigma$ and $\bar{K}N$ channels, 
the $W$ dependence of $d\sigma/(dtdW)$ is clearly different between 
$\pi^+ p \to K^{*+} (\mx\bx)^+$ (Fig.~\ref{fig:dsdtdw-all-plus}) and $\pi^- p \to K^{*0} (\mx\bx)^0$ (Fig.~\ref{fig:dsdtdw-all})
due to the absence of the $I=0$ $\Lambda$ resonance contributions in the former.
In fact, in the $\mx\bx= (\pi\Sigma)^+$ and $\bar{K}^0 p$ channels, 
we do not observe any of the following features: 
the huge bump observed below the $\bar{K}N$ threshold in the 
$\mx\bx=(\pi\Sigma)^0$ channels\footnote{The notation $(\pi\Sigma)^0$ collectively refers to the $\pi^+\Sigma^-$, $\pi^0\Sigma^0$, and $\pi^-\Sigma^+$ channels.} 
that originates from the two low-lying $1/2^-$ $\Lambda$ resonances; 
the sharp peak observed at $W\approx 1.52$~GeV in the $\mx\bx=(\pi\Sigma)^0$ and $(\bar{K}N)^0$ channels produced by the $\Lambda(1520)3/2^-$; 
or the bump observed around $W\approx 1.82$~GeV in the $\mx\bx=(\pi\Sigma)^0$ and $(\bar{K}N)^0$ channels 
that originates from the $\Lambda$ resonances in this energy region, particularly the $\Lambda(1820)5/2^+$.
On the other hand, in the $\mx\bx = (\pi\Sigma)^+$ channels of Model~B, a peak is observed at $W\approx 1.38$~GeV due to the $\Sigma(1385)3/2^+$, 
which is obscured by the existence of the two low-lying $1/2^-$ $\Lambda$ resonances in the $\mx\bx=(\pi\Sigma)^0$ channels.
Furthermore, in the $\mx\bx = \bar{K}^0 p$ channel, a bump is observed around $W\approx 1.78$~GeV, 
which is similar to the behavior in the $\mx\bx=\pi\Lambda$ channels but is absent in the $(\bar{K}N)^0$ channels.
This bump likely originates from the $\Sigma$ resonances in this energy region, particularly the $\Sigma(1775)5/2^-$, 
and is considered to be obscured by the $I=0$ contributions in the $(\bar{K}N)^0$ channels.

For the $\mx\bx=K\Xi$ channels, the $W$ dependence of $d\sigma/(dtdW)$ for $\mx\bx = K^+\Xi^0$ is noticeably smoother 
compared to that of the $\mx\bx=(K\Xi)^0$ channels shown in Fig.~\ref{fig:dsdtdw-all}. 
This would be attributable to the absence of $\Lambda$ resonances around $W=1.8$~GeV for the $\mx=K^+\Xi^0$ case.

Regarding the contributions of the $\pex$-exchange and $\vex$-exchange processes to $d\sigma/(dtdW)$ for the $\pi^+ p \to K^{*+} (\mx\bx)^+$ reactions, 
the $\pex$-exchange processes dominate the cross section below the $\bar{K}N$ threshold, 
while the contribution of the $\vex$-exchange processes can be comparable to, or even larger than, 
the $\pex$-exchange contribution.
This behavior is essentially the same as that observed in the $\pi^- p \to K^{*0} (\mx\bx)^0$ reactions.
The major differences between Model~A and Model~B in the $W$ dependence of $d\sigma/(dtdW)$ above the $\bar{K}N$ threshold 
seem to stem mostly from the differences in the $\vex$-exchange contribution.

\subsubsection{Partial-wave contributions}

\begin{figure}[tb]
\centering
\includegraphics[clip,width=0.48\textwidth]{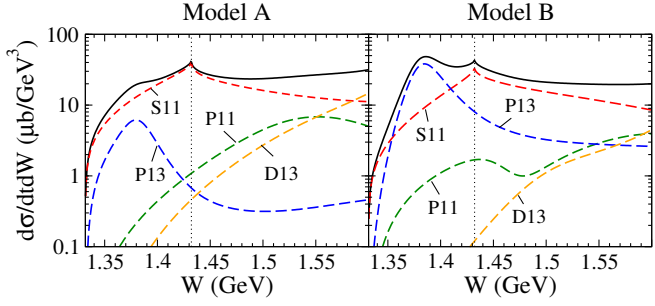}
\caption{
Contributions of individual partial waves in the $\mex \bi \to \mx\bx$ subprocess 
to $d\sigma/(dtdW)$ for the $\pi^+ p \to \mf^+ (\pi\Sigma)^+$ reaction at $|\moms{p}_\mi^\lfr| = 10$~GeV/$c$ and $t'=0$.
Only partial-wave contributions with $J \leq 3/2$ are shown for the $W$ range from the threshold up to $W=1.6$~GeV.
The left (right) panel shows the results for Model~A (Model~B).
The solid curves represent the full results, while the dashed curves correspond to those obtained 
by retaining only the specific single partial wave indicated in the figure when constructing the invariant amplitude for the $\mex \bi \to \mx\bx$ subprocess.
The vertical dotted line indicates the $\bar K N$ threshold.
}
\label{fig:dsdtdw-pwa-piS-plus}
\end{figure}
\begin{figure}[tb]
\centering
\includegraphics[clip,width=0.48\textwidth]{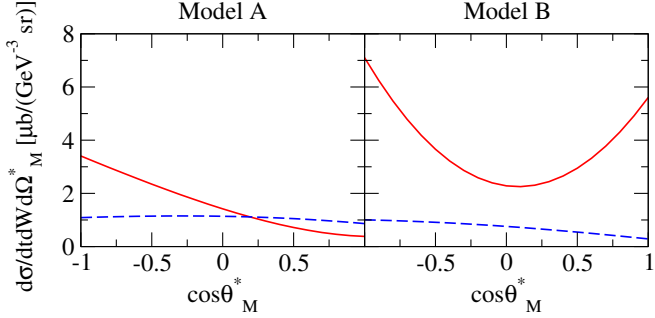}
\caption{
The $\cos\theta^*_\mx$ dependence of $d\sigma/(dtdWd\Omega^*_\mx)$ 
for the $\pi^+ p \to \mf^+ (\pi \Sigma)^+$ reaction at $|\moms{p}_\mi^\lfr| = 10$~GeV/$c$, $t'=0$, and $W=1381$~MeV. 
The azimuthal angle of the meson $\mx$ ($\phi^*_\mx$) is set to zero.
The left (right) panel shows the results for Model~A (Model~B).
The red solid curves represent the full results, while the blue dashed curves 
show the results obtained by excluding the contribution of the $P_{13}$ partial wave in the $\mex \bi \to \mx \bx$ subprocess.
}
\label{fig:dsdtdwdO-piS-plus-1381mev}
\end{figure}
\begin{figure}[tb]
\centering
\includegraphics[clip,width=0.48\textwidth]{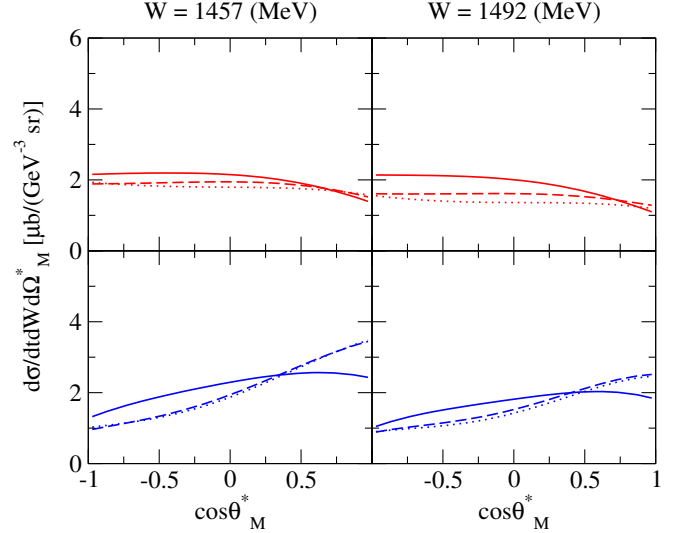}
\caption{
The $\cos\theta^*_\mx$ dependence of $d\sigma/(dtdWd\Omega^*_\mx)$ 
for the $\pi^+ p \to \mf^+ (\pi \Sigma)^+$ reaction 
at $|\moms{p}_\mi^\lfr| = 10$~GeV/$c$, $t'=0$. 
The results at $W=1457$ and $1492$~MeV are presented.
The azimuthal angle of the meson $\mx$ ($\phi^*_\mx$) is set to zero.
The top (bottom) panels show the results of Model~A (Model~B).
The solid curves represent the full results, while the dashed (dotted) curves 
show the results obtained by excluding the contribution of the $P_{11}$ partial wave
(the contributions of the $P_{11}$ and $D_{13}$ partial waves)
in the $\mex \bi \to \mx \bx$ subprocess.
}
\label{fig:dsdtdwdO-piS-plus-1457-1492mev}
\end{figure}

We now examine the partial-wave contributions of the $\mex \bi \to \mx\bx$ subprocess to $d\sigma/(dtdW)$ 
for the $\pi^+ p \to \mf^+ (\pi \Sigma)^+$ reaction.
As already mentioned, because the magnitudes of cross sections for the $\pi^+ p \to \mf^+ \pi^+ \Lambda$ 
reaction are exactly twice those for the $\pi^- p \to \mf^0 \pi^0 \Lambda$ reaction 
within our current model based on the $t$-channel one-meson exchange mechanism under isospin symmetry, 
the findings and consequences derived from the $\pi^- p \to \mf^0 \pi^0 \Lambda$ reaction apply equally to the $\pi^+ p \to \mf^+ \pi^+ \Lambda$ reaction.
Therefore, we do not present the separate results for the $\pi^+ p \to \mf^+ \pi^+ \Lambda$ reaction.

Figure~\ref{fig:dsdtdw-pwa-piS-plus} shows the contributions of individual partial waves in the $\mex \bi \to \mx\bx$ subprocess to $d\sigma/(dtdW)$ for the $\pi^+ p \to \mf^+ (\pi \Sigma)^+$ reactions.
In these reactions, only the $I=1$ partial waves contribute, unlike the case of the $(\pi\Sigma)^0$ channels.
The relative behavior among the $I=1$ partial-wave contributions is found to be quite similar to those observed in the $MB=\pi^\pm\Sigma^\mp$ channels (Fig.~\ref{fig:dsdtdw-pwa-piS-0c}).
In the $MB=\pi^\pm\Sigma^\mp$ channels, these $I=1$ contributions are subdominant and obscured by the $I=0$ partial waves; 
however, in the $(\pi\Sigma)^+$ channels, they are clearly manifested in the cross section due to the absence of the $I=0$ contributions.
Below the $\bar{K}N$ threshold, the $S_{11}$ and $P_{13}$ waves are the dominant contributors to the cross section.
Comparing Model~A and Model~B, while there is no significant difference in the magnitude of the $S_{11}$ contribution, 
the $P_{13}$ contribution in Model~B is approximately five times larger than that in Model~A around $W=1.38$~GeV.
This marked difference results in the clear peak structure observed in the cross section around $W=1.38$~GeV exclusively in Model~B.
This difference may be attributed to a variation in the coupling strength to the $\pi\Sigma$ channel for the $\Sigma(1381)3/2^+$ resonance, 
which is present in both models, or to the $\Sigma(1360)3/2^+$ state found in Model~B.
In the low-lying $Y^*$ region above the $\bar{K}N$ threshold, the $S_{11}$ contribution continues to be dominant, 
while the contributions from other partial waves remain very small. 
This behavior stands in sharp contrast to the $MB = \pi\Lambda$ channels, 
where the $P_{11}$ and $D_{13}$ contributions, both of which host low-lying $\Sigma^*$ resonances in Model~B, 
are comparable in magnitude to that of the $S_{11}$ wave.

\subsubsection{Angular distributions}

Figure~\ref{fig:dsdtdwdO-piS-plus-1381mev} shows the $\cos\theta^*_\mx$ dependence of $d\sigma/(dtdWd\Omega^*_\mx)$ 
for the $\pi^+ p \to \mf^+ (\pi\Sigma)^+$ reactions at $|\moms{p}_\mi^\lfr| = 10$~GeV/$c$, $t'=0$, and $W=1381$~MeV.
It can be seen that the angular dependence is significantly different between Model~A and Model~B, and 
this difference originates from the $P_{13}$ contribution in the $\mex \bi \to \mx \bx$ subprocess.
Similar to the discussion regarding Fig.~\ref{fig:dsdtdwdO-piL-1381mev}, this difference in the angular dependence can be understood as follows.
The $\cos\theta_\mx^*$ dependence of the cross section can qualitatively take the form:
$|{\cal A}_{S_{11}}|^2 + ({\cal A}_{S_{11}}{\cal A}_{P_{13}}^* + \textrm{c.c.})\cos\theta_\mx^* + |{\cal A}_{P_{13}}|^2 (3\cos^2\theta_\mx^* +1) +\cdots$,
where ${\cal A}_{S_{11}}$ (${\cal A}_{P_{13}}$) generically denotes the quantity encompassing the amplitude and other relevant factors associated with the $S_{11}$ ($P_{13}$) contribution 
of the $\pex$-exchange process, with its specific $\cos\theta_\mx^*$ dependence factored out in each respective term, and 
the minor terms involving either the $\vex$-exchange process or any partial wave other than $S_{11}$ and $P_{13}$ are suppressed.
In Model~A, where the $P_{13}$ contribution is smaller than the $S_{11}$ contribution, 
the term containing $|{\cal A}_{P_{13}}|^2$ has a much smaller effect than the terms 
containing $|{\cal A}_{S_{11}}|^2$ and $({\cal A}_{S_{11}}{\cal A}_{P_{13}}^* + \textrm{c.c.})$,
resulting in a linear behavior with respect to $\cos\theta_\mx^*$.
On the other hand, in Model~B, where the $P_{13}$ contribution is dominant, 
the term containing $|{\cal A}_{P_{13}}|^2$ dictates the cross section, 
leading to an upward-opening parabolic behavior.
As in the case of the $MB=\pi\Lambda$ channels, 
if precise data on the angular distribution for the $MB=(\pi\Sigma)^+$ channels become available, 
it may be possible to investigate the potential existence of another low-lying $3/2^+$ $\Sigma$ resonance suggested by our DCC models.

Figure~\ref{fig:dsdtdwdO-piS-plus-1457-1492mev} shows the $\cos\theta^*_\mx$ dependence of $d\sigma/(dtdWd\Omega^*_\mx)$ 
for the $\pi^+ p \to \mf^+ (\pi\Sigma)^+$ reactions at $|\moms{p}_\mi^\lfr| = 10$~GeV/$c$ and $t'=0$.
The results at $W=1457$ and $1492$~MeV are presented.
The former (latter) $W$ value corresponds to the real part of the pole mass of the low-lying $1/2^+$ ($3/2^-$) $\Sigma$ resonance found in Model~B.
As already mentioned in the discussion of Fig.~\ref{fig:dsdtdw-pwa-piS-plus}, 
in this $W$ region for these reactions, 
the $S_{11}$ contribution in the $\mex N \to \mx \bx$ subprocess is dominant, 
and the contributions from other partial waves are very small.
Consequently, the drastic $W$ dependence of the angular distribution caused by the $P_{11}$ and $D_{13}$ contributions,
which is observed exclusively in Model~B as seen in the $MB=\pi\Lambda$ channels, 
does not manifest in the $MB=(\pi\Sigma)^+$ channels.
As a result, no dramatic difference in the angular distribution is observed between Model~A and Model~B.

\subsection{$d\sigma/dt$ and $d\sigma/(dtdWd\phi^*_\mx)$}

\begin{figure*}[tb]
\centering
\includegraphics[clip,width=0.8\textwidth]{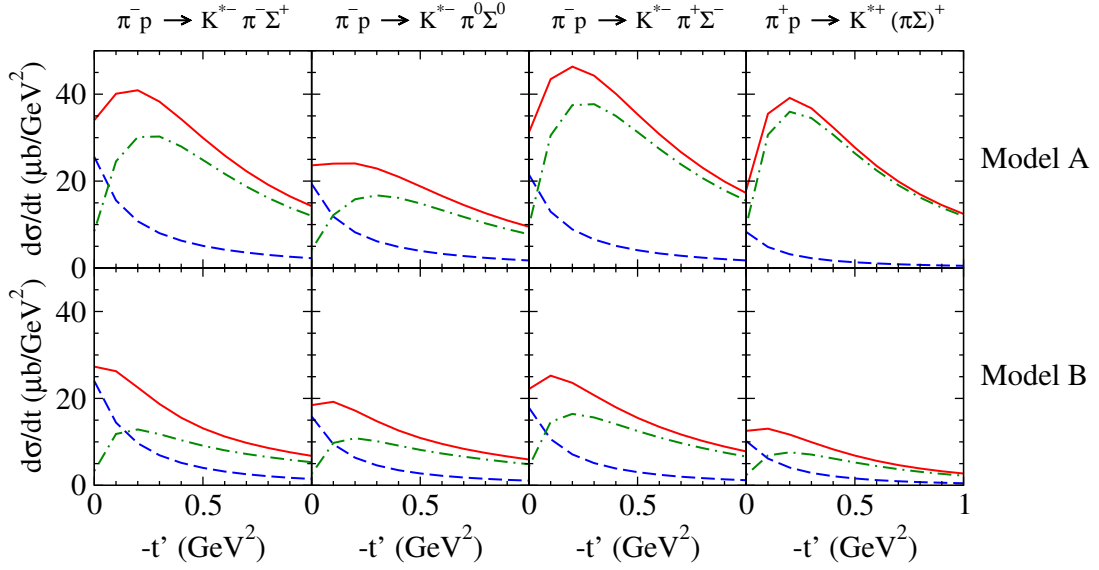}
\caption{
Differential cross section $d\sigma/dt$ for $\mi p \to K^* \pi \Sigma$ at $|\moms{p}_\mi^\lfr|=10$~GeV/$c$.
The results are obtained by integrating $W$ in the range from the $\pi\Sigma$ threshold to $W=2.1$~GeV.
The top (bottom) panels show the results of Model~A (Model~B).
The red solid curves represent the full results, while the dashed (dot-dashed) curves denote the results
where only the $\pex$-exchange ($\vex$-exchange) process is taken into account.
}
\label{fig:dsdt-10gev-piS}
\end{figure*}
\begin{figure*}[tb]
\centering
\includegraphics[clip,width=0.8\textwidth]{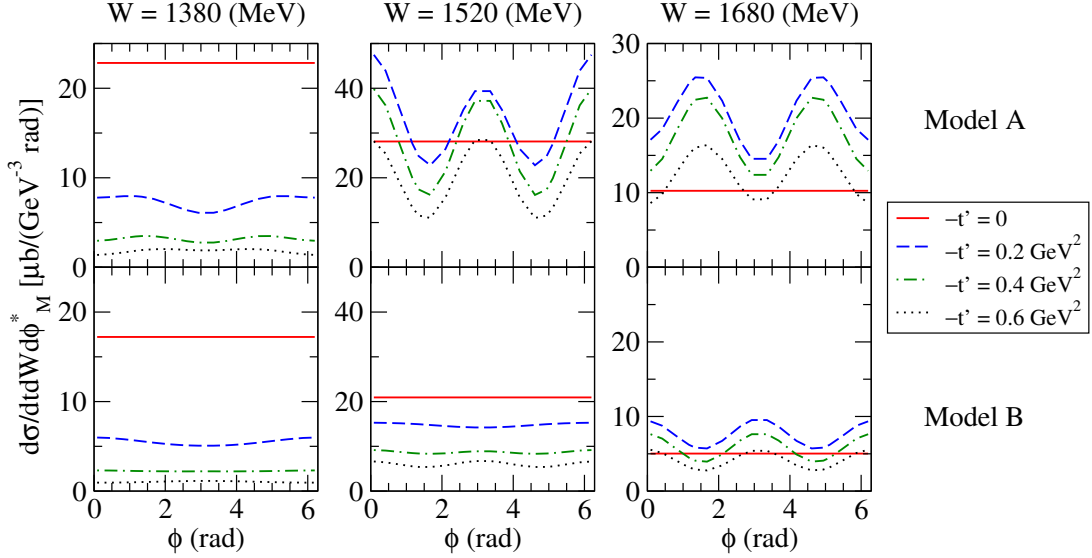}
\caption{
The $\phi^*_\mx$ dependence of $d\sigma/(dtdWd\phi^*_\mx)$ 
for the $\pi^- p \to \mf^- \pi^- \Sigma^+$ reaction 
at $|\moms{p}_\mi^\lfr| = 10$~GeV/$c$. 
The results at $W=1380$, $1520$, and $1680$~MeV are presented.
The top (bottom) panels show the results of Model~A (Model~B).
The red solid, blue dashed, green dot-dashed, and black dotted curves represent the results at $-t'=0$, $0.2$, $0.4$, and $0.6$~GeV$^2$, respectively.
}
\label{fig:dsdtdwdphi-piS}
\end{figure*}

Finally, we discuss the $t'$ dependence of $d\sigma/dt$ and the $\phi^*_\mx$ dependence of $d\sigma/(dtdWd\phi^*_\mx)$ at $|\moms{p}_\mi^\lfr| = 10$~GeV/$c$.
In general, the $\phi^*_\mx$ dependence arises at nonzero $t'$ values.
In our model, when considering only the $\pex$-exchange processes, 
the $\phi^*_\mx$ dependence does not appear even at nonzero $t'$ values; 
instead, it emerges from the $\vex$-exchange processes and the interference between the $\pex$- and $\vex$-exchange processes.\footnote{
From Eqs.~(\ref{eq:amp-pex})--(\ref{eq:amp-vexNmb}), 
and noting that $\moms{\hat p}_\pex^*$ and $\moms{\hat p}_\vex^*$ are parallel to the $z$ axis in the $\mx\bx$ center-of-mass frame, 
it can be seen that the amplitudes for the $\pex$- and $\vex$-exchange processes have a $\phi^*_\mx$ dependence of the form 
$\exp[i(S_\bi^z-S_\bx^z)\phi^*_\mx]{\cal M}_{\pex}'$ and $\exp[i(S_\bi^z-S_\bx^z)\phi^*_\mx]\{\sum_{S_\vex^z}\exp(iS_\vex^z\phi^*_\mx){\cal M}_{\vex(S_\vex^z)}'\}$, 
respectively, where ${\cal M}_{\mex}'$ represents the terms independent of $\phi_\mx^*$ in the amplitudes.
}
Therefore, the $\phi^*_\mx$ dependence of $d\sigma/(dtdWd\phi^*_\mx)$ 
provides valuable information regarding the relative magnitudes and 
phases of the amplitudes for the $\pex$- and $\vex$-exchange processes.
Consequently, like the $t'$ dependence of $d\sigma/dt$, it is expected to offer useful constraints for determining parameters 
such as $\Lambda_\pex$ and $\Lambda_\vex$ included in the high-energy part of the $\mi N \to \mf\mx\bx$ reaction, 
which is independent of the $\mex N \to \mx\bx$ subprocess.

Figure~\ref{fig:dsdt-10gev-piS} shows $d\sigma/dt$ for the $\pi p \to \mf \pi\Sigma$ reactions at $|\moms{p}_\mi^\lfr|=10$~GeV/$c$ up to $-t' \leq 1$~GeV$^2$.
The results are obtained by integrating $W$ over the range from the $\pi\Sigma$ threshold to $W=2.1$~GeV.
For these reactions, the $\pex$-exchange process is larger than the $\vex$-exchange process near $-t' = 0$.
The contribution of the $\pex$-exchange process decreases rapidly as $-t'$ increases, 
while that of the $\vex$-exchange process increases up to $-t' \approx 0.2$~GeV$^2$ 
and subsequently decreases more gradually than the $\pex$-exchange process.
The relative magnitudes of the $\pex$- and $\vex$-exchange contributions crossover around $-t' \approx 0.1$~GeV$^2$.
In the region $0.1 < -t' < 1$~GeV$^2$, the $\vex$-exchange contribution can be several to ten times larger than the $\pex$-exchange contribution, 
depending on the reaction channel.
Comparing Model~A and Model~B, no dramatic differences are observed in the magnitude or the $-t'$ dependence of the $\pex$-exchange process.
In contrast, a significant difference is found for the $\vex$-exchange process, particularly in its magnitude.
This is likely because the parameters associated with the quasi-two-body $\bar{K}^* N$ channel are only indirectly determined in Model~A and Model~B, 
which were obtained through a comprehensive analysis of the $\bar K N \to \bar K N, \pi\Sigma, \pi\Lambda, \eta\Lambda, \text{and } K\Xi$ reactions.
Thus, the marked difference can be attributed to the uncertainties in these parameters.
The $\mi N \to \mf\mx\bx$ reactions are expected to provide valuable information for the determination of parameters related to the quasi-two-body $\bar{K}^* N$ channel.

Figure~\ref{fig:dsdtdwdphi-piS} shows the $\phi^*_\mx$ dependence of $d\sigma/(dtdWd\phi^*_\mx)$ for the $\pi^- p \to \mf^- \pi^- \Sigma^+$ reaction 
at $|\moms{p}_\mi^\lfr| = 10$~GeV/$c$.
The results are presented at $W=1380$, $1520$, and $1680$~MeV.
As a general trend, it is observed that the $\phi^*_\mx$ dependence becomes more pronounced at higher $W$ values.
This is consistent with the fact that the contribution of the $\vex$-exchange process is generally larger at higher $W$, 
as shown in Figs.~\ref{fig:dsdtdw-all} and~\ref{fig:dsdtdw-all-plus}.

\section{\label{sec:summary}Summary}

In this work, we have developed a model for investigating the forward $K^*$ production reaction off the nucleon induced by a high-momentum $\pi$ beam, $\mi N \to \mf \mx \bx$, 
as an effective tool for establishing low-lying $Y^*$ resonances located below and just above the $\bar{K}N$ threshold.
Because conventional $K^- p$ scattering experiments cannot directly access the subthreshold region, 
and it is practically difficult to perform complete experiments just above the $\bar{K}N$ threshold, 
the proposed reaction provides a valuable complement for $Y^*$ spectroscopy.
In the considered high-energy peripheral production kinematics, we modeled the reaction based on $t$-channel one-meson exchange processes.
The half-off-shell scattering amplitudes for the $\mex \bi \to \mx\bx$ subprocesses were provided by the ANL-Osaka DCC model for the $S=-1$ sector, 
which provides two distinct parameter sets, Model~A and Model~B, derived from our 2014 ANL-Osaka DCC analysis.

Our 2014 DCC models extracted potential new resonance states below the $\bar{K}N$ threshold, 
most notably a new $3/2^+ \Sigma$ resonance in addition to the well-established $\Sigma(1385)3/2^+$ state.
Other poorly established new states, such as a low-lying $1/2^- \Sigma$ (in Model~A) and a $1/2^+ \Lambda$ (in Model~B), 
were also identified in addition to two $1/2^-$ $\Lambda$ resonances.
Using these models, we evaluated how the predicted low-lying $Y^*$ resonances manifest in various observables.

The predicted differential cross sections $d\sigma/(dtdW)$ demonstrate significant enhancements in the subthreshold region for channels such as $\pi\Lambda$ and $\pi\Sigma$.
In particular, Model~A predicts a distinct peak in the $\pi\Lambda$ invariant mass spectrum at $W \approx 1.31$~GeV and a second peak at $W \approx 1.38$~GeV, 
stemming from the new $3/2^+ \Sigma$ state and the well-established $\Sigma(1385)3/2^+$, respectively.
In Model~B, however, the new resonance and the $\Sigma(1385)3/2^+$ appear as a single peak due to their closely located poles.
It is known from previous experiments that a distinct peak around $W \approx 1.31$~GeV does not exist, 
rendering the situation in Model~A unrealistic. 
However, as demonstrated in Model~B, two $3/2^+$ $\Sigma$ resonances could constitute a single peak in the $\pi\Lambda$ invariant mass spectrum, 
much like how two $1/2^-$ $\Lambda$ resonances form a single peak in the $\pi\Sigma$ spectrum. 
In the latter case, the possibility that two $3/2^+$ $\Sigma$ resonances exist below the $\bar{K}N$ threshold cannot be ruled out at present.

We have also demonstrated that when invariant mass distributions alone 
fail to distinguish overlapping resonances, 
angular distributions [$d\sigma/(dtdW d\Omega_\mx^*)$] offer strong discriminatory power.
Interference effects between dominant and subdominant partial waves (e.g., $S_{01}$ and $P_{13}$) lead to distinct angular behaviors, 
such as clear linear and/or parabolic dependencies on $\cos \theta_\mx^*$, 
which vary dramatically between Model~A and Model~B.

Furthermore, we investigated the $t'$ dependence of $d\sigma/dt$ and the $\phi_\mx^*$ dependence of $d\sigma/(dtdWd\phi_\mx^*)$.
The investigation of these dependencies provides useful information for constraining the relative magnitudes and phases of the $\pex$- and $\vex$-exchange processes. 
This, in turn, helps determine model parameters associated with the high-energy $\pi \to \mf \mex$ transition process and the propagation of the exchanged $\mex$ mesons, 
which are not determined by the DCC model employed to describe the $\mex N \to \mx\bx$ subprocesses.

In conclusion, the $\mi N \to \mf \mx \bx$ reaction offers highly sensitive observables for disentangling the complex $Y^*$ mass spectrum 
below and just above the $\bar{K}N$ threshold.
The kinematic conditions required for this reaction are expected to be accessible at modern hadron facilities such as J-PARC.We hope that 
the theoretical predictions presented in this work will strongly motivate future high-statistics measurements of differential cross sections and angular distributions, 
thereby advancing our understanding of low-lying $Y^*$ resonances.
Ultimately, performing a simultaneous comprehensive analysis that incorporates both the $\mi N \to \mf \mx \bx$ and $\bar{K}N$ reactions 
will completely unravel the $Y^*$ mass spectrum in the $S=-1$ sector.

\begin{acknowledgments}
The authors are grateful to Prof. Hiroyuki Noumi for providing the J-PARC E31 data and for helpful information regarding the J-PARC experiments.
T.-S.H.L. is supported by the U.S. Department of Energy, Office of Science, Office of Nuclear Physics, under Contract No. DE-AC02-06CH11357.
\end{acknowledgments}

\appendix
\section{\label{app:weight}Model for the weight function ${\cal W}(\mmf)$}

In order to evaluate the weight function ${\cal W}(\mmf)$ defined in Eq.~(\ref{eq:weight}), 
which contains information on the propagation of the outgoing $\mf$ meson 
and its decay into the $\mfo\mft$ pair,
we employ a separable potential model for the amplitude of the $J^P=1^-$ and $I=1/2$
partial wave of the $\pi K$ scattering.
This model was originally developed in Ref.~\cite{Kamano2014} 
to construct a Green function for the quasi-two-body $\bar K^* N$ states.

In this model, the $\mf$ propagator $D_\mf$ and the $\mf \to \mfo \mft$ decay amplitude
$\tilde{\cal M}_{\mfo\mft,\mf}(|\moms{\kappa}|)$ contained in ${\cal W}(\mmf)$ are expressed as
\begin{align} 
D_\mf = 
&
 \frac{1}{2m_\mf^0}\tau_\mf(M_\mf),
\end{align} 
and
\begin{align} 
i\tilde {\cal M}_{\mfo\mft, \mf} (|\moms{\kappa}|) =
& 
 -\frac{i}{(2\pi)^3} 
\nonumber\\
&\times
 \sqrt{(2\pi)^3 2E_\mfo(|\moms{\kappa}|)} 
 \sqrt{(2\pi)^3 2E_\mft(|\moms{\kappa}|)} 
\nonumber\\
&\times
 \sqrt{(2\pi)^3 2m_\mf^0}
 f_{\mfo\mft,\mf}(|\moms{\kappa}|) ,
\end{align}
where $\tau_\mf(E)$ and the vertex function $f_{\mfo\mft,\mf}(l)$ are defined by
\begin{align}
\tau_\mf (E) =  
&
 \frac{1}{E - m_\mf^0 -\Sigma_\mf(E)},
\\
\Sigma_\mf (E) =  
&
 \int l^2 dl \frac{f_{\mf,\mfo\mft}(l) f_{\mfo\mft,\mf}(l)}{E-E_\mfo(l) - E_\mft(l) + i\varepsilon},
\\
f_{\mfo\mft,\mf}(l) =
&
 \frac{\bar g_{\mfo\mft\mf}}{\sqrt{m_\pi}}
 \left(\frac{l}{m_\pi}\right)
 \left(\frac{\Lambda^2}{\Lambda^2 + l^2}\right).
\end{align}

\begin{figure}[t]
\centering
\includegraphics[clip,width=0.48\textwidth]{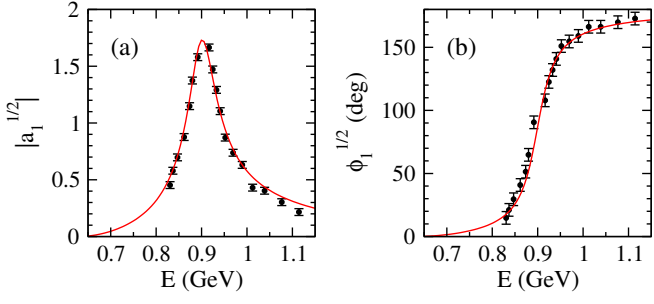}
\caption{
The $I=1/2$ and $L=1$ partial-wave amplitude of $\pi \bar K$ scattering.
The data are taken from Ref.~\cite{Aston1988}.
The definitions of the amplitude $a^{1/2}_1$ and its phase $\phi^{1/2}_1$ 
are given in Eqs.~(4.4) and~(4.5) of Ref.~\cite{Aston1988}.
}
\label{fig:pik}
\end{figure}

Here, the pion and kaon masses are taken to be $m_\pi = 138.5$~MeV and $m_K = 493.7$~MeV, 
while the values of the parameters associated with this model are given by
$m_\mf^0 = 930.35$~MeV, $\bar g_{\mfo\mft\mf} =-0.15243$, and 
$\Lambda = (0.57885~\mathrm{fm})^{-1}$.
The parameters were determined by fitting to the 
$I=1/2$ and $L=1$ partial-wave amplitude for $\pi \bar K$ scattering extracted in Ref.~\cite{Aston1988}. 
The fit results are presented in Fig.~\ref{fig:pik}, which demonstrates that
this model describes the $I=1/2$ and $L=1$ partial-wave amplitude well up to $E=1.15$~GeV.
With the parameters obtained from the fit, this model yields 
a $K^*$ pole mass of $(899.34 -i29.69)$~MeV.
Figure~\ref{fig:weight} shows the $M_\mf$ dependence of the weight function ${\cal W}(M_\mf)$.
The weight function exhibits a clear peak at $M_\mf \approx 0.9$~GeV as expected.

\section{\label{sec:vpp-vvp-vtx}Model for the $\mi \to \mf \mex$ vertices}

In this Appendix, we summarize the model used for evaluating the invariant amplitudes for the $\mi \to \mf \mex$ vertices,
which is based on the phenomenological SU(3) Lagrangians given in Ref.~\cite{Kamano2014}.

\subsection{Model for the $\mi \to \mf \pex$ vertex}
The invariant amplitude for the $\mi \to \mf \pex$ vertex is given by
\begin{align}
&
 i{\cal M}_{\mf(\lambda_\mf,I^z_\mf) \pex(I^z_\pex), \mi(I^z_\mi)}(p_\mf,p_\pex;p_\mi) = 
\nonumber\\
&
 \bra{\mf(p_\mf;\lambda_\mf,I^z_\mf)\pex(p_\pex;I^z_\pex)} iL_{VPP'} \ket{\mi(p_\mi;I^z_\mi)}.
\end{align}
For the Lagrangian describing the interaction between one octet vector meson and two octet pseudoscalar mesons, $L_{VPP'}$, 
we use the following form~\cite{Kamano2014}:
\begin{align}
L_{VPP'} =
&
 L^{\rm L}_{VPP'} \times L^{\rm F}_{VPP'} ,
\end{align}
with
\begin{align}
L^{\textrm{L}}_{VPP'} =
& 
 iP(\partial_\mu P')V^\mu ,
\\
L^{\textrm{F}}_{VPP'} =
& 
 g''[[P_{\bm{8}} \otimes P'_{\bm{8}}]^{(\bm{8}_2)}\otimes V_{\bm{8}}]^{(\bm{1})} .
\end{align}
Here, $L^{\rm L}_{VPP'}$ [$L^{\textrm{F}}_{VPP'}$] represents
the Lorentz [SU(3) flavor] part of the $VPP'$ Lagrangian.

The matrix elements of the Lorentz and SU(3) flavor parts for the $\mi \to \mf \pex$ vertex are then given\footnote{
See, e.g., Ref.~\cite{DeSwart} for calculations associated with the SU(3) algebra.
} by
\begin{align}
\bra{\mf(p_\mf,\lambda_\mf)\pex(p_\pex)} L^{\rm L}_{VPP'} \ket{\mi(p_\mi)} = 
&
 i(-ip_\mi^\mu)[\varepsilon^{(\lambda_\mf)}_{\mf\mu}]^* ,
\label{eq:vpp-l1}
\end{align}
\begin{align}
\bra{\mf\pex} L^{\rm F}_{VPP'} \ket{\mi} =
&
 -\frac{1}{\sqrt{2}}\left(-\frac{g''}{\sqrt{24}}\right) 
\nonumber\\
&\times
 \cg{I_\mf I^z_\mf}{I_{\pex}I^z_{\pex}}{I_{\mi}I^z_{\mi}} ,
\label{eq:vpp-f1}
\end{align}
for the case where the operator $P'$ ($P$) in the Lagrangian contracts with $\mi$ ($\pex$), and by
\begin{align}
\bra{\mf(p_\mf,\lambda_\mf)\pex(p_\pex)} L^{\rm L}_{VPP'} \ket{\mi(p_\mi)} =
&
 i(+i p_\pex^\mu)[\varepsilon^{(\lambda_\mf)}_{\mf\mu}]^* ,
\label{eq:vpp-l2}
\end{align}
\begin{align}
\bra{\mf\pex} L^{\rm F}_{VPP'} \ket{\mi} =
&
 +\frac{1}{\sqrt{2}}\left(-\frac{g''}{\sqrt{24}}\right) 
\nonumber\\
&\times
 \cg{I_\mf I^z_\mf}{I_{\pex}I^z_{\pex}}{I_{\mi}I^z_{\mi}} ,
\label{eq:vpp-f2}
\end{align}
for the case that the operator $P'$ ($P$) in the Lagrangian contracts with $\pex$ ($\mi$).
Using Eqs.~(\ref{eq:vpp-l1})-(\ref{eq:vpp-f2}), we have
\begin{align}
&
 \bra{\mf(p_\mf;\lambda_\mf,I^z_\mf)\pex(p_\pex;I^z_\pex)} iL_{VPP'} \ket{\mi(p_\mi;I^z_\mi)}  =
\nonumber\\
&\qquad
 i \cg{I_\mf I^z_{\mf}}{I_\pex I^z_\pex}{I_\mi I^z_\mi} 
 \left(-\frac{1}{\sqrt{2}}\right)\left(-\frac{g''}{\sqrt{24}}\right)
\nonumber\\
&\qquad
\times
 (p_\mi+p_\pex)_\mu [\varepsilon^{(\lambda_\mf)\mu}_{\mf}]^*  .
\end{align}

\subsection{Model for the $\mi \to \mf \vex$ vertex}
The invariant amplitude for the $\mi \to \mf \vex$ vertex is given by
\begin{widetext}
\begin{align}
i{\cal M}_{\mf(\lambda_\mf,I^z_\mf) \vex(S^z_\vex,I^z_\vex), \mi(I^z_\mi)}(p_\mf,p_\vex;p_\mi) =
&
 \bra{\mf(p_\mf;\lambda_\mf,I^z_\mf)\vex(p_\vex;S^z_\vex,I^z_\vex)} iL_{VV'P} \ket{\mi(p_\mi;I^z_\mi)}  .
\end{align}
\end{widetext}
For the Lagrangian describing the interaction between two octet vector mesons and one octet pseudoscalar meson, $L_{VV'P}$, 
we use the following form (the model framework is based on that of Ref.~\cite{Kamano2014}, but the $VV'P$ Lagrangian is presented here for the first time):
\begin{equation}
L_{VV'P} = L^{\rm L}_{VV'P} \times L^{\rm F}_{VV'P} ,
\end{equation}
with
\begin{align}
L^{\rm L}_{VV'P} =
& 
 \epsilon^{\alpha\beta\gamma\delta} (\partial_\alpha V_\beta) (\partial_\gamma V'_\delta) P ,
\\
L^{\rm F}_{VV'P} =
& 
 g'''[[V_{\bm{8}} \otimes V'_{\bm{8}}]^{(\bm{8}_1)}\otimes P_{\bm{8}}]^{(\bm{1})} .
\end{align}
Here, $L^{\rm L}_{VV'P}$ [$L^{\textrm{F}}_{VV'P}$] represents
the Lorentz [SU(3) flavor] part of the $VV'P$ Lagrangian,
and $\epsilon^{0123} \equiv +1$.

The matrix elements of the Lorentz and SU(3) flavor parts for the $\mi \to \mf \vex$ vertex are then given by
\begin{align}
&
 \bra{\mf(p_\mf,\lambda_\mf), \vex(p_\vex,S_{\vex}^z)} L^{\rm L}_{VV'P} \ket{\mi(p_\mi)} =
\nonumber\\
&
 \epsilon^{\alpha\beta\gamma\delta} 
 (+ip_{\mf\alpha})[\varepsilon^{(\lambda_{\mf})}_{V\beta}]^*
 (+ip_{\vex\gamma})[\varepsilon^{(S^z_{\vex})}_{\vex\delta}]^* ,
\label{eq:vvp-l1}
\end{align}
\begin{align}
\bra{\mf\vex} L^{\rm F}_{VV'P} \ket{\mi} =
&
 -\frac{3}{2\sqrt{2}}\left(-\frac{g'''}{\sqrt{30}}\right)
\nonumber\\
&\times 
 \cg{I_\mf I^z_\mf}{I_{\vex}I^z_{\vex}}{I_\mi I^z_\mi} ,
\label{eq:vvp-f1}
\end{align}
for the case where the operator $V$ ($V'$) in the Lagrangian contracts with $\mf$ ($\vex$), and by
\begin{align}
&
 \bra{\mf(p_\mf,\lambda_\mf), \vex(p_\vex,S_{\vex}^z)} L^{\rm L}_{VV'P} \ket{\mi(p_\mi)} =
\nonumber\\
&\qquad
 \epsilon^{\alpha\beta\gamma\delta} 
 (+ip_{\vex\alpha})[\varepsilon^{(S^z_{\vex})}_{\vex\beta}]^*
 (+ip_{\mf\gamma})[\varepsilon^{(\lambda_\mf)}_{\mf\delta}]^* ,
\label{eq:vvp-l2}
\end{align}
\begin{align}
\bra{\mf\vex} L^{\rm F}_{VV'P} \ket{\mi} =
&
 -\frac{3}{2\sqrt{2}}\left(-\frac{g'''}{\sqrt{30}}\right)
\nonumber\\
&
 \times \cg{I_\mf I^z_\mf}{I_{\vex}I^z_{\vex}}{I_\mi I^z_\mi} ,
\label{eq:vvp-f2}
\end{align}
for the case where the operator $V$ ($V'$) in the Lagrangian contracts with $\vex$ ($\mf$).
Using Eqs.~(\ref{eq:vvp-l1})-(\ref{eq:vvp-f2}), we have
\begin{widetext}
\begin{align}
&
\bra{\mf(p_\mf;\lambda_\mf,I^z_\mf)\vex(p_\vex;S^z_\vex,I^z_\vex)} iL_{PVV'} \ket{\mi(p_\mi;I^z_\mi)} =
\nonumber\\
&
\qquad 
 i \cg{I_\mf I^z_{\mf}}{I_\vex I^z_\vex}{I_\mi I^z_\mi}
 \left(-\frac{3}{\sqrt{2}}\right)\left(-\frac{g'''}{\sqrt{30}}\right)
 \epsilon^{\alpha\beta\gamma\delta} 
 (+ip_{\mf\alpha})[\varepsilon^{(\lambda_\mf)}_{\mf\beta}]^*
 (+ip_{\vex\gamma})[\varepsilon^{(S^z_{\vex})}_{\vex\delta}]^* .
\end{align}
\end{widetext}

\bibliography{ref}

\end{document}